\documentclass{article}[14pt]
\usepackage{blindtext}
\usepackage{geometry}
\usepackage[utf8]{inputenc}
\usepackage[T1]{fontenc}

\usepackage[english]{babel}
\usepackage{authblk}
\usepackage{cancel}

\usepackage{amsmath}
\usepackage{amssymb}
\usepackage{amsfonts}
\usepackage{graphicx}
\usepackage{comment}
\usepackage{float}
\usepackage{xcolor}
\usepackage[normalem]{ulem}
\usepackage{caption}
\usepackage{array}
\usepackage{setspace}
\usepackage{subcaption}
\usepackage{booktabs}
\usepackage{xcolor}
\usepackage[tableposition=above]{caption}
\usepackage{soul}
\usepackage{pifont}
\usepackage{adjustbox}
\usepackage{multirow}
\usepackage{abstract}

\usepackage{hyperref}
\usepackage{cleveref}

\usepackage{csquotes}
\usepackage[style=authoryear,natbib=true,maxbibnames=99,doi=false]{biblatex}
\renewbibmacro{in:}{}
\DeclareNameAlias{sortname}{family-given}

\begin{document}

\singlespacing
\font\myfontA=cmr12 at 16pt
\font\myfontB=cmr12 at 12pt
\title{\myfontA Bayesian Longitudinal Tensor Response Regression for Modeling Neuroplasticity  \\ \vspace{.25cm}}

\author{\myfontB Suprateek Kundu$^1$, Alec Reinhardt$^1$\footnote{{\it Corresponding author}: Alec Reinhardt, Department of Biostatistics, UT MD Anderson Cancer Center, 1400 Pressler St., Floor 4, FCT4.6000, Houston, TX 77030. {\it E-mail}: aereinhardt@mdanderson.org.}, Serena Song$^2$, Joo Han$^2$, M. Lawson Meadows$^2$, Bruce Crosson$^3$, and Venkatagiri Krishnamurthy$^{2,4}$ \\
{$^1$Department of Biostatistics, UT MD Anderson Cancer Center} \\
{$^2$Center for Visual and Neurocognitive Rehabilitation, Atlanta Veterans Affairs Medical Center} \\
{$^3$Departments of Neurology \& Imaging and Radiological Sciences, Emory University} \\
{$^4$Division of Geriatrics and Gerontology, Emory University}
}

\date{\today}
\maketitle


{\centerline{\bf{Abstract}}}

A major interest in longitudinal neuroimaging studies involve investigating voxel-level neuroplasticity changes due to treatment and other factors across visits. To assess such changes, voxel-wise analysis is routinely used. However they are beset with several pitfalls, which can compromise the accuracy of these approaches. We propose a novel Bayesian tensor response regression approach for longitudinal imaging data, which is able to pool information across spatially-distributed voxels in order to infer significant neuroplasticity changes adjusting for covariate effects. The proposed method leads to massive dimension reduction via a low-rank decomposition, while preserving the spatial configurations of the voxels when estimating coefficients. It also enables feature selection via a joint credible regions approach that respects the shape of the posterior distribution for more accurate inference. In addition to group-level inferences, the proposed method is also able to infer significant brain changes at the individual-level that are highly desirable for studying personalized disease or recovery  trajectories. An efficient Markov chain Monte Carlo (MCMC) sampling scheme is developed for implementing the proposed method. The numerical advantages of the proposed approach in terms of prediction and feature selection compared to the routinely used voxel-wise regression are highlighted via extensive simulation studies. Subsequently, we apply the approach to a longitudinal aphasia dataset to study neuroplasticity changes measured from task functional MRI data collected from a group of subjects who were administered either a {\color{black} control} intervention or an  intention treatment at baseline, and were followed up over  subsequent visits.  Our analysis revealed that while the control therapy showed long-term increases in brain function/activity, the intention treatment produced predominantly short-term activity changes, which were concentrated in a number of localised regions depending on the treatment. In contrast, the standard voxel-wise regression failed to detect any significant neuroplasticity changes after multiplicity adjustments, which is biologically implausible and implies lack of power.

\vskip 5pt


\section{Introduction}

In the event of stroke, highly interconnected neural systems are disrupted due to cell death in the core lesioned brain areas, cell dysfunction in the perilesional brain areas, and compromised activity in remote brain areas due to hypometabolism, neurovascular uncoupling, and aberrant neurotransmission \citep*{pekna2012modulation}. These neurobiological changes are expected to result in considerable neuroplasticity in the stroke brain, defined as the phenomenon of spontaneous or treatment-enhanced restoration and re-organization of brain functioning that supports relearning of lost functions
(\citealp*{pekna2012modulation}; \citealp{crosson2019neuroplasticity}; \citealp{Reid2016}). Aphasia is a stroke-related acquired language impairment disorder characterized by brain lesions, which has been widely studied in literature \citep{watila2015factors}. One of the key aspects of aphasia is that it is possible to design behavioral interventions that can result in clinically meaningful language gains even during the chronic phase that are potentially governed by the principles of neural plasticity \citep{Cappa2000,Wilson2021}. 
The disease severity and overall disease prognosis, as well as neural plasticity and associated recovery, may depend on the type of aphasia, which is partially characterized by the location and size of brain lesions \citep{crosson2019neuroplasticity}. Given that aphasia outcomes have heterogeneity across neural regions, time, and subjects, there is a growing need to better understand how to effectively apply targeted clinical interventions to improve outcomes. This type of approach may require one to go beyond the routinely used group-level analysis to predict personalized neural plasticity changes that account for heterogeneity. 

Functional magnetic resonance imaging (fMRI) techniques for investigating neural plasticity changes in aphasia have been around for two decades and are particularly appealing in terms of allowing researchers to investigate functional changes in the brain after stroke. However, the findings from these studies have been highly variable. Some studies have supported a role for the right hemisphere \citep{Blank2002, Crinion2005, Turkeltaub2012}, while others have reinforced the importance of residual left hemisphere language areas \citep{Saur2006, Griffis2017, Fridriksson2012}. Most recently, several studies have suggested that domain-general networks may play a role in supporting recovery from aphasia \citep{Brownsett2014, Geranmayeh2014}. Researchers generally agree that all of these types of mechanisms are likely to play some role in recovery from aphasia, and that the relative importance of different mechanisms probably depends on the location and extent of the lesions, the phase of recovery, and other clinical characteristics. However, there is often no consensus as to which specific regions are more likely to be activated in post-stroke aphasia (PSA) compared to healthy individuals, and this scenario is further exacerbated by typically small sample sizes in stroke studies. 
Such variability in findings in aphasia literature is potentially due to inherent heterogeneity between samples, which is often overlooked by current approaches relying on group-level comparisons. Existing methods essentially tend to ignore the heterogeneity within each treatment group, which may arise either spontaneously or from clinical, demographic, or other characteristics. Hence, there is a critical unmet need of developing robust statistical approaches for mapping personalized neuroplasticity trajectories from heterogeneous samples in aphasia studies that go beyond simple group-level comparisons. 

Standard analytical methods in aphasia literature routinely use a voxel-wise approach \citep{naylor2014voxelwise,benjamin2014behavioral} that performs the analysis independently for each voxel. Unfortunately, this approach has several pitfalls that are often overlooked. First, a voxel-wise analysis is only able to include those samples that do not have a lesion present at a given voxel, resulting in a loss of effective sample size and power coupled with unreliable estimation. This is particularly not desirable in the presence of moderate to small samples routinely encountered in aphasia and other brain lesion studies. Second, the total number of model parameters in voxel-wise regression models increases linearly with the number of voxels, which typically ranges from a few thousand to close to a hundred thousand in neuroimaging applications. The resulting lack of parsimony in voxel-wise methods results in overfitting issues. Lastly, a voxel-wise analysis is unable to respect the spatial configurations of the voxels, nor is it able to pool information across neighboring voxels. Thus, it essentially treats voxels as independently distributed, and it ignores the fact that functional imaging data usually involve simultaneous signal changes in spatially distributed brain regions \citep{van2010exploring}. 
{\color{black}Moreover, these voxel-wise methods require stringent} multiplicity adjustments to account for a large number of hypothesis tests \citep{eklund2016cluster}. 
{\color{black}Classical multiplicity adjustments such as the Bonferroni method often result in overly conservative estimates and do not have any mechanism to account for the correlations between neighboring voxels. A more common solution is to impose spatial clustering in the voxel-wise significance maps via Cluster-extent inference \citep{chumbley2010topological}. Typically, this method first fits the model voxel-wise and subsequently performs multiplicity corrections at the level of clusters of voxels. Other methods such as the one proposed in \cite{park2021permutation} {\color{black}first compute voxel-level test statistics that are combined within clusters and scaled by a spatial covariance matrix (obtained via a permutation approach) to derive a global test statistic to be used for inference.} While such {\color{black}heuristic} approaches are useful, they may still result in inadequate voxel-level coefficient estimation, and inferential results are not guaranteed to be optimal. }

In this article, we develop a powerful alternative to voxel-wise methods by proposing a novel longitudinal Bayesian tensor response regression {\color{black}(l-BTRR)} model for longitudinal brain imaging studies that overcomes the aforementioned challenges. The proposed approach treats the brain image as a tensor-valued outcome, which is regressed on covariates via low-rank coefficient matrices involving voxel-specific effects that respect the spatial configuration of voxels. The low-rank structure results in massive dimension reduction, resulting in model parsimony that is critically important for high-dimensional neuroimaging applications involving tens of thousands of voxels. It is also designed to borrow information across neighboring voxels to estimate effect sizes, which results in increased accuracy and precision for coefficient estimation. Another desirable feature is the ability of the Bayesian tensor model to impute outcome values for missing voxels where needed. These features are advantageous compared to a univariate voxel-wise analysis or an alternate multivariable regression approach that simply assigns a unique regression coefficient corresponding to each voxel. Moreover, the Bayesian framework naturally enables one to infer significant covariate effects as well as neuroplasticity changes over visits, and also report measures of uncertainty. This is achieved via Bayesian joint credible regions that respect the shape of the posterior distribution and provide improvements over the routinely used coordinate-wise credible interval approach. The proposed approach is implemented via an efficient posterior computation scheme involving Markov chain Monte Carlo (MCMC) developed in this paper. 

The proposed method differentiates between group-level effects that may be either time-invariant or time-varying and are learned by pooling information across samples, from individual-specific effects that define unique characteristics specific to a given subject. By accommodating subject-specific effects, the proposed approach allows us to infer personalized neuroplasticity trajectories over visits that are not feasible under the routinely used group-level comparisons. Incorporating heterogeneity is critical for our aphasia neuroimaging application comprising a group of post-stroke subjects at varying stages of aphasia, who were randomly assigned to either the intention therapy \citep{crosson2008intention} or the {\color{black} control} therapy group at baseline, and were followed up over 3 months post-intervention. {\color{black} At the group-level, coefficient estimation and} feature selection {\color{black} are} also much improved in our aphasia data analysis, {\color{black} such that} the proposed approach infers several clusters of brain regions with significant neuroplasticity changes. In contrast, the  voxel-wise regression analysis for this dataset is unable to infer any group-level significant neuroplasticity changes after multiplicity adjustments. This seems biologically implausible and underlines the limitations of voxel-wise analysis. {\color{black}To validate the operating characteristics of the proposed model, we additionally conduct extensive simulation studies, {\color{black} where} the proposed approach {\color{black} is} compared to voxel-wise regression methods and cross-sectional tensor models.}

Although motivated by stroke literature, the proposed approach can be applied to a wide range of longitudinal neuroimaging studies where it is of interest to regress the brain image on covariates to infer significantly associated voxels. 
 To our knowledge, this work is one of the first Bayesian tensor-based methodology developed for high-dimensional longitudinal neuroimaging applications involving heterogeneous samples that go beyond the limited and fairly recent literature on cross-sectional {\color{black}tensor models that is summarised here.} {\color{black} In the frequentist paradigm, \cite{rabusseau2016higher} constructed a regression model with a tensor response exploiting a low-rank structure, but without inducing sparsity that is often required to identify important tensor nodes and cells. \cite{li2017parsimonious} proposed an envelope-based tensor response model relying on a generalized sparsity principle that is designed to identify linear combinations of the response irrelevant to the regression.  More recently, \cite{sun2017store} developed a new class of models, referred to as STORE, that impose element-wise sparsity in tensor coefficients. While useful, frequentist approaches are unable to perform inference required for feature selection and can not quantify uncertainty, both of which are naturally possible under Bayesian methods. In the Bayesian paradigm, \cite{Guhaniyogi2021} proposed a Bayesian tensor response regression approach that is built upon a multi-way stick-breaking shrinkage prior on the tensor coefficients. \cite{spencer2019bayesian} further generalized the approach by \cite{Guhaniyogi2021} to jointly identify
activated brain regions due to a task and connectivity between different brain regions. In a recent paper, \cite{guha2021bayesian} expand on previous work to develop a generalized Bayesian linear modeling framework with a symmetric tensor response and scalar predictors. Unfortunately, none of the above existing approaches cater to the scenario of longitudinal imaging studies, which presents challenges due to subject- and voxel-specific trajectories of neuroplasticity {\color{black} and is the focus of this article.}

The rest of the article is organized as follows. Section 2.1 provides a primer on tensor-based modeling, Section 2.2 develops the longitudinal Bayesian tensor response regression modeling approach, Section 2.3 develops a novel feature selection strategy using Bayesian joint credible regions, and Section 2.4 outlines the method to infer neuroplasticity maps from the proposed model. Section 3 develops the MCMC steps for efficient posterior computation scheme. Section 4 describes extensive simulation studies that compare the performance of the proposed method with state-of-the-art competing approaches, and Section 5 reports the results from the aphasia data analysis.

\section{Methods}
\subsection{Overview of tensor models} 




One of the earliest proposed techniques for tensor modeling is known as  PARAFAC decomposition \citep{kolda2009tensor}, which is a special case of the more general Tucker decomposition \citep{kolda2009tensor}, and will be used for our purposes throughout the article.  In mathematical terms, the mode $D$ tensor denoted as $\mathcal{X}_R$ (and belonging to the space $\mathbb{R}^{p_1\times \ldots \times p_D}$) may be expressed as the sum of $R$ independent outer products of rank-1 tensor margins. In particular, one may write $\mathcal{X}_R = \sum_{r=1}^R \lambda_r x_{1\bullet,r} \circ x_{2\bullet,r} \circ \ldots \circ x_{D\bullet,r},$
\noindent where `$\circ$' denotes an outer product, the set of vectors $\{x_{d\bullet,r} \in \mathbb{R}^{p_d \times 1}\}_{d=1}^D$ are the so-called tensor margins having rank 1, for any given $\{r=1,\ldots,R \}$, $\lambda_r$ represents the weight for the $r$th channel, and $R$ represents the chosen rank of the tensor. In practice, we often have $D=3$ for three-dimensional images in our applications. Given the above definition, 
the $(j_1,\ldots,j_D)$-th element of the rank-R tensor $\mathcal{X}_R$ can be expressed as $
{\mathcal{X}_R}_{j_1,j_2,\ldots,j_D} = \sum_{r=1}^R \lambda_r x_{1\bullet,r,j_1} x_{2\bullet,r,j_2} \ldots x_{D\bullet,r,j_D}$. We note that the tensor margins $\{x_{d\bullet,r}\}_{d=1}^D$ are only identifiable up to a permutation and multiplicative constant, unless some additional constraints are imposed. However the lack of identifiability of tensor margins does not pose any issues for our setting, since the tensor product is fully identifiable, which is sufficient for our primary goal of coefficient estimation. Hence we do not impose any additional identifiability conditions on the tensor margins, which is consistent with Bayesian tensor modeling literature \citep{Guhaniyogi2020}. The tensor decomposition is visually illustrated in Figure \ref{fig:Figure1} for the three-dimensional case. 

\begin{figure}[H]
  \begin{subfigure}[t]{.55\textwidth}
    \centering
    \includegraphics[width=\linewidth]{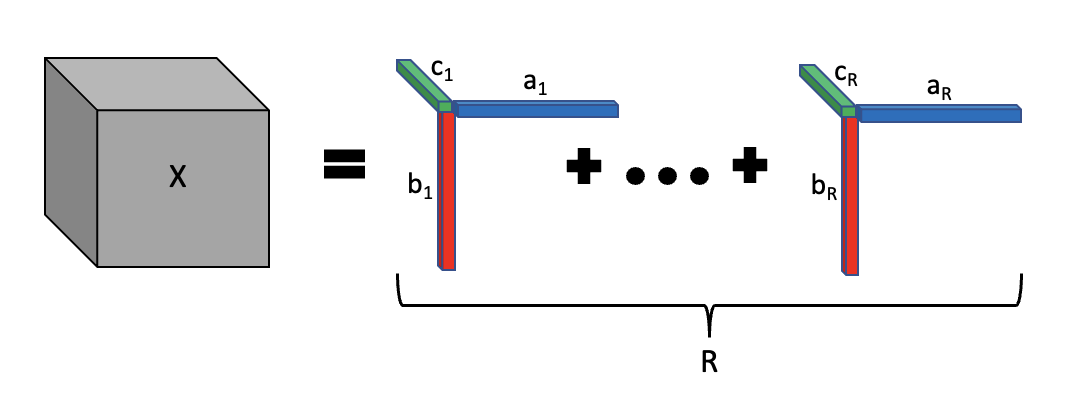}
  \end{subfigure}
  \hfill
  \begin{subfigure}[t]{.42\textwidth}
    \centering
    \includegraphics[width=\linewidth]{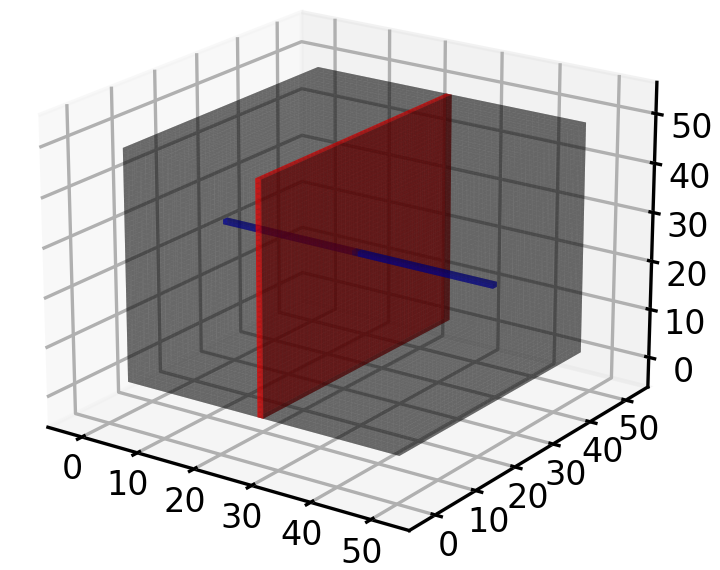}
  \end{subfigure}
  \begin{subfigure}[t]{.42\textwidth}
    \centering
    \includegraphics[width=\linewidth]{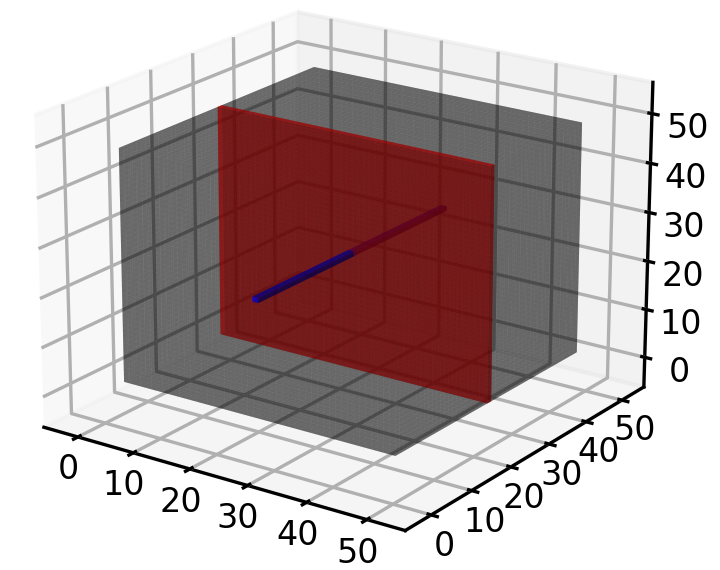}
  \end{subfigure}
  \hfill
  \begin{subfigure}[t]{.42\textwidth}
    \centering
    \includegraphics[width=\linewidth]{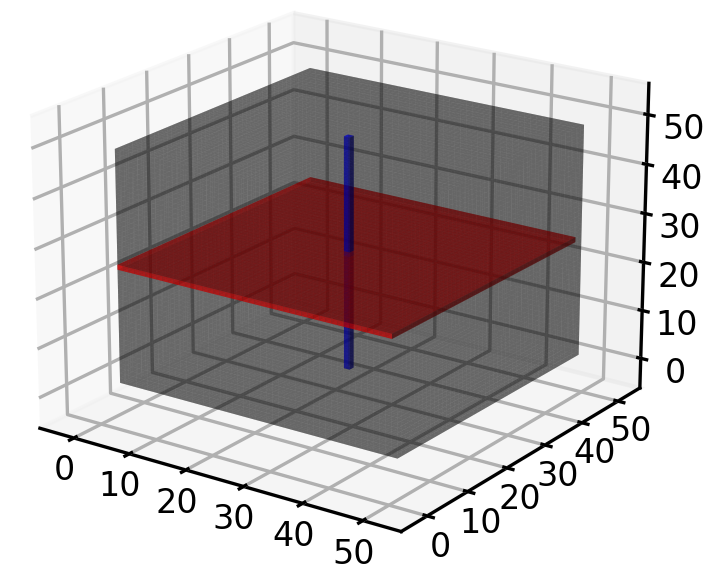}
  \end{subfigure}
  \caption{\textbf{Tensor visualization.} Top left panel provides a graphic of a rank-$R$ tensor decomposition for a 3-dimensional tensor $X$, represented as the sum of tensor products of vectors $a_r$, $b_r$, and $c_r$, $1\leq r \leq R$. The remaining panels illustrate tensor slices (red) and fibers (blue) corresponding to each of the 3 dimensions of a 3-way tensor cube.}
  \label{fig:Figure1}
\end{figure}

In addition to the tensor margins, other lower dimensional objects are naturally embedded within a tensor structure. These include tensor fibers that can be visualized as a thin thread of points generated when varying only one of the tensor modes, while keeping the other modes fixed. For example for a three-way tensor ($D=3$), mode-1 fibers correspond to the collection of $d_1$-dimensional vectors defined as $\{\mathcal{X}_{R,\cdot,j_2,j_3} = (\mathcal{X}_{R,1,j_2,j_3},\ldots,\mathcal{X}_{R,d_1,j_2,j_3})^T:1\le j_2\le d_2, 1\le j_3\le d_3 \}$. Mode-2 and Mode-3 fibers can be defined similarly. On the other hand, tensor slices are defined as lower dimensional sub-spaces of a tensor that are generated by varying two tensor modes simultaneously, while keeping the third tensor mode fixed (for the $D=3$ case). For example, the $(1,2)$ tensor slice corresponding to the point $j^*$ on tensor mode-3 may be represented as a collection $\{  \mathcal{X}_{R,j_1,j_2,j^*} : 1\le j_1\le d_1, 1\le j_2 \le d_2 \}$, where $j^* \in \{1,\ldots, d_3 \}.$ The tensor fibers and slices are illustrated in Figure \ref{fig:Figure1}, and these structures will be useful for understanding different aspects of the proposed model. More importantly, tensor slices will be directly instrumental for estimating the tensor margins even in the presence of missing voxels, as outlined in the following section.

\subsection{Proposed Model}


{\noindent \underline{Notations:}} Consider the observed three-dimensional brain image $\mathcal{Y}_{ti}\in \mathbb{R}^{p_1\times p_2\times p_3}$ for the $i$-th sample ($i=1,\ldots,n,$) and the $t$-th visit, with $p_1,p_2,p_3,$ voxels along the three dimensions, and $t=0,..,T$ corresponding to the baseline and $T$ follow-up visits. Our method can accommodate varying numbers of visits across subjects in a relatively straightforward manner. The brain image can represent the chosen measure of brain activity or structure as determined by the specific application, and it is treated as a tensor object in our article. Let us denote the measurement corresponding to the $v$-th voxel of $\mathcal{Y}_{ti}$ as $y_{ti}(v)$, where $v\in \mathcal{V}$ {\color{black},} {\color{black} with $\mathcal{V}$ denoting the space of all voxels within the brain mask}. Further denote the set of covariates for the $i$-th sample as $({\bf c}_i,{\bf x}_i, {\color{black}{\bf z}_{ti}})$, where ${\bf c}_i$ $(M\times 1)$ and ${\bf x}_i (S\times 1)$ induce time-varying and time-invariant effects of the outcome respectively, {\color{black} and ${\bf z}_{ti}$ $(Q\times 1)$ induce effects of time-varying covariates.} We further denote $\mathcal{T}_{ti}$ as the time duration between baseline and visit $t$ for the $i$th sample (assumed to be zero at baseline).


\vskip 10pt

{\noindent \underline{Longitudinal Tensor Response Regression Model:}} We propose the following generic longitudinal tensor response regression model for three-dimensional images ($D=3$):

\begin{eqnarray}
&&\mathcal{Y}_{ti} = \mathcal{M} + B_i  + \Gamma\times \mathcal{T}_{ti} + \Theta_i \times \mathcal{T}_{ti}+ 
\sum_{m=1}^M\mathcal{B}_{tm} c_{im} + \sum_{s=1}^S \mathcal{D}_s x_{is} + {\color{black} \sum_{q=1}^Q \mathcal{C}_{q} z_{tiq}} + \mathcal{E}_{ti}, \quad \label{eq:longBTRR} 
\end{eqnarray}

\noindent where $\times$ denotes matrix product, $\mathcal{M}$ represents the population-level intercept term that can be assigned a tensor structure as $\mathcal{M} = \sum_{r=1}^R \mu_{1\bullet,r} \circ  \mu_{2\bullet,r} \circ  \mu_{3\bullet,r}$,  $\Gamma = \sum_{r=1}^R \gamma_{1\bullet,r} \circ \gamma_{2\bullet,r} \circ \gamma_{3\bullet,r}$ represents the population-level regression slopes corresponding to the follow-up times ($\mathcal{T}$), 
$\mathcal{B}_{tm} = \sum_{r=1}^R \beta_{1\bullet,rtm} \circ \beta_{2\bullet,rtm} \circ \beta_{3\bullet,rtm} $ denotes the regression coefficients at visit $t$ corresponding to the time-varying effects of covariate $c_m$ ($m=1,\ldots,M$) that capture longitudinal changes, and the time-invariant effects are given as $\mathcal{D}_{s} = \sum_{r=1}^R \delta_{1\bullet,rs} \circ \delta_{2\bullet,rs} \circ  \delta_{3\bullet,rs} $ ($s=1,\ldots,S$) {\color{black} and $\mathcal{C}_{q} = \sum_{r=1}^R \chi_{1\bullet,rq} \circ \chi_{2\bullet,rq} \circ \chi_{3\bullet,rq}$. Note that $\mathcal{D}_s$ corresponds to time-invariant covariates ${\bf x}_i$, while $\mathcal{C}_q$ corresponds to time-varying covariates ${\bf z}_{ti}$.} While these effects capture population-level changes, model (\ref{eq:longBTRR}) also includes subject-specific random intercept term $B_i = \sum_{r=1}^R b_{1\bullet,ri} \circ  b_{2\bullet,ri} \circ  b_{3\bullet,ri}$  that captures baseline deviations for samples and a subject-specific time slope $ \Theta_i=\sum_{r=1}^R \theta_{1\bullet,ri} \circ \theta_{2\bullet,ri} \circ \theta_{3\bullet,ri} $ that captures variations in the longitudinal trajectory across samples. Both the terms $(B_i,\Theta_i)$ capture individual-level variations and are important for accommodating heterogeneity that is an important consideration in our neuroplasticity problems. Finally, 
$\mathcal{E}_{ti}\in \mathbb{R}^{p_1\times p_2\times p_3}$ denotes the random residual errors that are assumed to be independently distributed and Gaussian. 
That is, we assume $\epsilon_{ti}(v)\sim N(0,\sigma_\epsilon^2)$ independently for all voxels $v \in \mathcal{V}$, which is similar to standard 
assumptions made in tensor modeling literature \citep{Guhaniyogi2021}.


We note that the proposed model can be interpreted as a non-trivial adaptation of routinely used linear mixed models in longitudinal studies to tensor-valued outcomes. Simultaneously, it may be also be considered an extension of the Bayesian tensor method in \citep{Guhaniyogi2021} proposed in the context of cross-sectional single-subject fMRI time-series data, to longitudinal neuroimaging studies involving multiple subjects. In addition to the longitudinal set-up, there are additional differences in the prior specifications on the tensor margins  as elaborated below.



\vskip 10pt

{\noindent \underline{Prior Specification:}}  
The prior specifications corresponding to the parameters in (\ref{eq:longBTRR}) are as follows:
\begin{eqnarray}
& \gamma_{d\bullet,r} \sim \mathcal{N}\left(\mathbf{0},\tau^{\gamma} \mathbf{W}^{\gamma}_{d,r}\right),  
\beta_{d\bullet,rtm} \sim \mathcal{N}\left(\mathbf{0},\tau^{\beta}_{tm} \mathbf{W}^{\beta}_{d,r}\right), \mbox{ } 
\delta_{d\bullet,rs} \sim \mathcal{N}\left(\mathbf{0},\tau^{\delta}_{s} \mathbf{W}^{\delta}_{d,r}\right), \mbox{ }
{\color{black}\chi_{d\bullet,rq} \sim \mathcal{N}\left(\mathbf{0},\tau^{\chi}_{s} \mathbf{W}^{\chi}_{d,r} \right)}, \nonumber \\
& \theta_{d\bullet,ri} \sim \mathcal{N}\left(\mathbf{0},\tau^{\theta}_{i} \mathbf{W}^{b}_{d,r}\right), \mbox{ } b_{d\bullet,ri} \sim \mathcal{N}\left(\mathbf{0},\tau^{b}_{i} \mathbf{W}^{b}_{d,r}\right),  \mbox{ }  
 \mu_{d\bullet,r} \sim \mathcal{N}\left(\mathbf{0},\tau^{\mu} \mathbf{W}^{\mu}_{d,r}\right), d=1,2,3, \label{eq:longBTRRprior}
\end{eqnarray}

\noindent where all global scale parameters in (\ref{eq:longBTRRprior}) follow $\tau \sim \text{Gamma}(a_{\tau},b_{\tau})$, all the residual variance terms follow an inverse Gamma prior, i.e.   $\sigma^2_{\epsilon,ti} \sim IG(a_\epsilon, b_\epsilon)$.  The covariance matrices $\mathbf{W}$ in (\ref{eq:longBTRRprior}) capture the correlations for each tensor margin, and are constructed to be positive semi-definite with dimension $p_d\times p_d$ corresponding to the $d$-th margin. In order to reduce the number of covariance parameters that can rapidly increase with the tensor dimensions, we propose a parametric low-rank structure on $\mathbf{W}$, which also has the advantage of spatial smoothing. In other words, the prior covariance structure encourages spatially contiguous clusters of voxels to have highly correlated activation coefficients.  We specify the following low-rank structure $\mathbf{W}^{\gamma}_{d,r} = \mbox{diag}\left(\sqrt{w^{\gamma}_{d,r,1}},\ldots, \sqrt{w^{\gamma}_{d,r,p_d}}\right) \times \Lambda_{d,r}^{\gamma} \times  \mbox{diag}\left(\sqrt{w^{\gamma}_{d,r,1}},\ldots, \sqrt{w^{\gamma}_{d,r,p_d}}\right)$, where the matrix $\Lambda_{d,r}^{\gamma}$ imposes spatial correlations such that $\Lambda_{d,r}^{\gamma}(k_1,k_2) = \exp\{-\alpha_{dr}^{\gamma}(k_1-k_2)^2 \}$ which translates to decreasing prior correlations with increasing distance between the $k_1$ and $k_2$th elements for the $d$-th margin ($k_1,k_2=1,\ldots,p_d$). The correlations also depend on the lengthscale or smoothness parameter $\alpha^{\gamma}_{dr}$ that is assigned a prior distribution $\alpha^{\gamma}_{dr} \sim \text{Gamma}(a_{\alpha},b_{\alpha})$, with higher $\alpha$ implying lower correlations. 
The diagonal variance terms in $\mathbf{W}^{\gamma}_{d,r}$ are expressed as $(w^{\gamma}_{d,r,1},\ldots, w^{\gamma}_{d,r,p_d})$, and these are assumed to be equal for simplicity, i.e. $w^{\gamma}_{d,r,1}=\ldots=w^{\gamma}_{d,r,p_d}=w^{\gamma}_{d,r}$. The following hierarchical priors are imposed on the unknown variance terms:
$w^{\gamma}_{d,r}\sim \text{Exp}(\lambda_{d,r}^{\gamma}/2), \ \lambda_{d,r}^{\gamma}\sim \text{Gamma}(a_{\lambda},b_{\lambda})$. Therefore, the $(k_1,k_2)$th off-diagonal element of $\mathbf{W}^{\gamma}_{d,r}$ are given as {\color{black}$w^{\gamma}_{d,r}\exp\{-\alpha_{dr}^{\gamma}(k_1-k_2)^2\}$}. The covariance matrices for the other regression coefficients in (\ref{eq:longBTRRprior}) are constructed similarly but understood to have distinct variance and length-scale parameters. A list of parameters in the model with a brief description is provided in Table \ref{tab:Table1}, where $\mathbb{R}$ denotes the set of real numbers and $\mathbb{N}$ is the set of natural numbers. 

\begin{table}[H]
    \centering
    \begin{tabular}{*3c}
        \toprule
        Parameter & Description & Range \\
        \midrule
        $(p_1, p_2, p_3)$ & Size of tensor dimensions for 3D tensor & $\mathbb{N}^3$ \\        
        $\mathcal{Y}_{ti}$ & AUC outcome tensor for visit $t$ and subject $i$ & $\mathbb{R}^{p_1\times p_2 \times p_3}$ \\
        $\mathcal{V}_i$ & Set of observed voxels corresponding to the $i$th subject & $\mathcal{V}_i \subset \mathcal{V}$ \\
        $\mathcal{T}_{ti}$ & Observed time of follow-up for $t$th visit and $i$th subject & $\mathbb{R}$ \\
        $\mathbf{c}_i$ & Covariates with time-varying effects for subject $i$th subject & $\mathbb{R}^{M \times 1}$ \\
        $\mathbf{x}_i$ & Covariates with time-invariant effects for subject $i$th subject & $\mathbb{R}^{S \times 1}$ \\
        $\mathbf{z}_{ti}$ & Time-varying covariates with time-invariant effects for $i$th subject and $t$th visit & $\mathbb{R}^{Q \times 1}$ \\
        $\mathcal{M}$ & Population-level intercept & $\mathbb{R}^{p_1 \times p_2 \times p_3}$ \\
        $B_i$ & Subject-specific intercept & $\mathbb{R}^{p_1\times p_2 \times p_3}$ \\
        $\Gamma$ & Population-level time slope & $\mathbb{R}^{p_1\times p_2 \times p_3}$ \\
        $\Theta_i$ & Subject-specific time slope & $\mathbb{R}^{p_1\times p_2 \times p_3}$ \\
        $\mathcal{B}_{tm}$ & Time-varying effect of m$th$ clinical covariate $\mathbf{c}_{i,m}$ & $\mathbb{R}^{p_1\times p_2 \times p_3}$ \\
        $\mathcal{D}_s$ & Time-invariant effect of s$th$ clinical covariate $\mathbf{x}_{i,s}$ & $\mathbb{R}^{p_1\times p_2 \times p_3}$ \\
        $\mathcal{C}_q$ & Time-invariant effect of $q$th time-varying covariate $\mathbf{z}_{tiq}$ & $\mathbb{R}^{p_1 \times p_2 \times p_3}$ \\
        $\mathcal{E}_{ti}$ & Random residual error &         $\mathbb{R}^{p_1\times p_2 \times p_3}$ \\
        $\sigma_{\epsilon,ti}^2$ & Variance of residual error (equal across voxels) & $\mathbb{R}^+$ \\
        
        $\mu_{d\bullet,r}$ & Tensor margin for $d$th dimension and $r$th channel for $\mathcal{M}$  & $\mathbb{R}^{p_d \times 1}$ \\
        $b_{d\bullet,ri}$ & Tensor margin for $d$th dimension and $r$th channel for $B_i$  & $\mathbb{R}^{p_d \times 1}$ \\
        $\gamma_{d\bullet,r}$ & Tensor margin for $d$th dimension and $r$th channel for $\Gamma$ & $\mathbb{R}^{p_d \times 1}$ \\
        $\theta_{d\bullet,ri}$ & Tensor margin for $d$th dimension and $r$th channel for $\Theta_i$ & $\mathbb{R}^{p_d \times 1}$ \\
        $\beta_{d\bullet,rtm}$ & Tensor margin for $d$th dimension and $r$th channel for $\mathcal{B}_{tm}$  & $\mathbb{R}^{p_d \times 1}$ \\
        $\delta_{d\bullet,rs}$ & Tensor margin for $d$th dimension and $r$th channel for $\mathcal{D}_s$  & $\mathbb{R}^{p_d \times 1}$ \\
        $\chi_{d\bullet,rq}$ & Tensor margin for $d$th dimension and $r$th channel for $\mathcal{C}_q$ & $\mathbb{R}^{p_d \times 1}$ \\
        $\tau$ & Global tensor margin variance scaling term & $\mathbb{R}^+$ \\
        $\mathbf{W}_{d,r}$ & Covariance matrix for $d$th tensor margin and $r$th channel & $\mathbb{R}^{p_d \times p_d}$ \\
        $\Lambda_{d,r}$ & AR-1 covariance matrix for $d$th tensor margin and $r$th channel & $\mathbb{R}^{p_d \times p_d}$ \\
        $\lambda_{d,r}$ & Rate parameter corresponding to diagonal elements of $\mathbf{W}_{d,r}$ & $\mathbb{R}^+$ \\
        $\alpha_{d,r}$ & Lengthscale parameter used to define $\Lambda_{d,r}$ & $\mathbb{R}^+$ \\
        \bottomrule
    \end{tabular}
    \caption{Summary of notations used in the longitudinal Bayesian tensor response regression model.}
    \label{tab:Table1}
\end{table}

The construction of the covariance matrices is expected to explicitly capture spatial correlations (in an unsupervised manner) corresponding to neighboring voxels, which is more flexible compared to specifying independence on the tensor margins as in \cite{Guhaniyogi2020}. The extent and strength of such correlations will depend on posterior distributions (see Section 3) that combines the likelihood resulting from model (\ref{eq:longBTRR}) with the priors in (\ref{eq:longBTRRprior}). The posterior distribution will be used for estimation under a Bayesian framework, resulting in data-adaptive correlation estimates that are allowed to vary over brain regions. 
In our experience, the resulting set-up yields model flexibility while simultaneously ensuring model parsimony. 
 Having specified the prior distributions, we now highlight the specific advantages under the tensor modeling framework below. 

\vskip 10pt

{\noindent \underline{Dimension Reduction via Tensor Structure:}} Consider the set of voxels {\color{black} in the brain mask} $\{v:v\in \mathcal{V} \}$ that is common across all subjects. At the voxel level, model (\ref{eq:longBTRR}) may be re-written as

\begin{eqnarray}
  y_{ti}(v) = \mathcal{M}(v) + B_i(v) +  \Gamma(v) \mathcal{T}_{ti} + \Theta_i(v) \mathcal{T}_{ti} +  \sum_{m=1}^M\mathcal{B}_{tm}(v) c_{i,m} + \sum_{s=1}^S \mathcal{D}_s(v) x_{is} + {\color{black}\sum_{q=1}^Q \mathcal{C}_q(v) z_{tiq}} + \epsilon_{ti}(v), \label{eq:vxl-longBTRR}
\end{eqnarray}

\noindent where $v {\color{black}\in \mathcal{V}}$ refers to the $v$th voxel. Instead of treating each voxel as a separate unit in (\ref{eq:vxl-longBTRR}), the voxel-specific coefficients are modeled using a low-rank PARAFAC decomposition \citep{rabanser2017introduction} that pools data across neighboring voxels to estimate a given voxel-specific coefficient.  For example, the coefficient  $\Gamma(k_1,k_2,k_3)$ is expressed as $\sum_{r=1}^R\gamma_{1k_1,r} \gamma_{2k_2,r} \gamma_{3k_3,r},$ where  $\gamma_{1\bullet,r}(p_1\times 1),\gamma_{2\bullet,r}(p_2\times 1)$ and $\gamma_{3\bullet,r}(p_3\times 1)$ denote the margins of the tensor corresponding to $\Gamma$, and $R$ denotes the rank of the PARAFAC decomposition that indicates the number of channels used for constructing the low-rank decomposition.  Under this low-rank PARAFAC decomposition, the number of distinct parameters for each regression coefficient matrix is given as $R(p_1+p_2+p_3)$ instead of a total of  $p_1p_2p_3$ distinct parameters. Such a tensor decomposition leads to a massive reduction in the number of parameters compared to existing methods, which makes the proposed approach computationally scalable to tens of thousands of voxels. Moreover,  model (\ref{eq:longBTRR}) may be generalized to include coefficient-specific ranks. A similar interpretation holds for the other tensor coefficients.

\vskip 10pt

{\noindent \underline{Preserving Spatial Configurations:}} Before fitting the tensor model, the voxels in the image are mapped to a regularly spaced grid that is more amenable to a tensor-based treatment. Such a mapping preserves the spatial configurations of the voxels that provides significant benefits over a univariate voxel-wise analysis or a multivariable analysis that vectorizes the voxels without regard for spatial configurations. Although the grid mapping may not preserve the exact spatial distances between voxels, this is of limited consequence in our experience, given the ability to capture correlations between neighboring elements in the tensor margins as elaborated previously. To better understand how spatial smoothing is induced between the regression coefficients for neighboring voxels, note that the tensor coefficients for $\Gamma$ corresponding to the neighboring voxels $(k_1,k_2,k_3)$ and $(k_1^*,k_2,k_3)$ for $k_1\ne k_1^*$ are given as $\Gamma(k_1,k_2,k_3)=\sum_{r=1}^R\gamma_{1k_1,r} \gamma_{2k_2,r} \gamma_{3k_3,r},$ and $\Gamma(k_1^*,k_2,k_3)=\sum_{r=1}^R\gamma_{1k_1^*,r} \gamma_{2k_2,r} \gamma_{3k_3,r},$ respectively. These coefficients share many common elements from the tensor margins $\gamma_{2,\bullet}$ and $\gamma_{3,\bullet}$ that induces correlations that is given by $Cov\big(\Gamma(k_1,k_2,k_3), \Gamma(k^*_1,k_2,k_3)\big)=\sum_{r=1}^R Cov\big(\gamma_{1k_1,r}, \gamma_{1k^*_1,r}\big)Var(\gamma_{2k_2,r})Var(\gamma_{3k_3,r})$.  Under the proposed prior in (\ref{eq:longBTRRprior}), we note that $Cov\big(\gamma_{1k_1,r}, \gamma_{1k^*_1,r}\big) = w^{\gamma}_{1,r}\exp\{-\alpha^{\gamma}_{1r} (k_1-k^*_1)^2 \}$, which implies \newline $Cov\big(\Gamma(k_1,k_2,k_3), \Gamma(k^*_1,k_2,k_3)\big)=\sum_{r=1}^R  w^{\gamma}_{1,r}w^{\gamma}_{2,r}w^{\gamma}_{3,r}\exp\{-\alpha^{\gamma}_{1r} (k_1-k^*_1)^2 \}$. This covariance decreases with the distance between the voxel indices $k_1,k^*_1$ for the first margin, thereby inducing spatial smoothing. 
Clearly, this is a desirable feature when one expects voxel activations that form spatially distributed clusters in different regions of the brain, as supported in literature \citep*{woo2014cluster}. 

\vskip 10pt

{\noindent \underline{Pooling information across voxels:}} A desirable feature of the tensor construction is that it is able to estimate voxel-specific coefficients using the information from neighboring voxels by estimating the tensor margins under the inherent low-rank structure. This feature yields more robust and reproducible brain maps illustrating significant voxels that are more robust to missing voxels and noise in the images. As a natural byproduct, this feature can be handily used for robust imputation of imaging outcomes corresponding to missing voxels, {\color{black} as} empirically validated via our extensive numerical studies in Section 4. In contrast, the ability to borrow information across neighboring voxels is completely missing in voxel-wise analysis, which treats the coefficients across voxels as independent without regard to their spatial configurations. 

For example, consider a toy scenario involving the estimation of the element $\Gamma(1,3,1)$ that is expressed as $\sum_{r=1}^R \gamma_{11,r}\gamma_{23,r}\gamma_{31,r}$ in model  (\ref{eq:longBTRR}). The estimation of coefficients proceeds through the estimation of the tensor margins $\{ (\gamma_{1\bullet,r},\gamma_{2\bullet,r},\gamma_{3\bullet,r}): r=1,\ldots,R \}$. We note that the elements $\{\gamma_{11,r}, r=1,\ldots,R\}$ are inherently contained in the tensor coefficients corresponding to all voxels in the tensor slice given by $\{(1,k_2,k_3), k_2=1,
\ldots,p_2,k_3=1,\ldots,p_3\}$ (refer to Figure \ref{fig:Figure1}). Similarly, the tensor margin elements $\{\gamma_{23,r},r=1,\ldots,R\}$ are contained in the  tensor coefficients corresponding to the tensor slice  $\{(k_1,3,k_3), k_1=1,\ldots,p_1,k_3=1,\ldots,p_3 \}$. A similar interpretation holds for the tensor margin elements $\{\gamma_{31,r},r=1,\ldots,R\}$ that can be learned using the voxels lying on the tensor slice $\{(k_1,k_2,1), k_1=1,\ldots,p_1,k_2=1,\ldots,p_2 \}$. Hence by pooling information across voxels contained in suitable tensor slices, the tensor margin parameters  $\{\gamma_{11,r},\gamma_{23,r},\gamma_{31,r}, r=1,\ldots,R \}$ can be learned in an effective manner that is more robust to  missing voxels as elaborated below.


\vskip 10pt

{\noindent \underline{Accommodating missing/redundant voxels:}} A desirable feature of the proposed model is that it is able to produce accurate results even when there is small to moderate proportion of missing voxels in the image. This is important for applications in stroke studies, where voxels lying in lesion areas show no activity and are considered missing. Consider the set of voxels $\mathcal{V}_i$ that is observed for the $i$th subject ($i=1,\ldots,n$). In the presence of subsets of missing or redundant voxels that may vary across individuals, the proposed model (\ref{eq:vxl-longBTRR}) may be modified {\color{black} by considering voxels $v \in \mathcal{V}_i$, where $\mathcal{V}_i$ denotes a subject-specific set of observed voxels.} Clearly, the only difference {\color{black} is that this model} assumes varying set of missing voxels across samples, instead of an identical set of missing voxels in (\ref{eq:vxl-longBTRR}). However, {\color{black} even with $v \in \mathcal{V}_i$}, the proposed model is still able to preserve its appealing features such as dimension reduction, accounting for spatial contiguity, and robust estimation in the presence of subsets of missing/redundant voxels. Since the tensor regression coefficients are expressed as a low-rank decomposition that involves outer products of tensor margins, each element in the tensor margin is learned by pooling information across corresponding subsets of tensor slices that comprise non-missing voxels. This facilitates the estimation of the voxel-specific coefficients corresponding to missing voxels in the image. Although there is some loss of information in not being able to use all the voxels on a tensor slice for estimating the tensor margins due to missingness, such loss is manageable when the proportion of missing voxels is not overly large (see simulation studies in Section 4).  When multiple samples are present with varying locations of missing voxels (as inevitably occurs in stroke samples due to the varying size, shape, and location of lesions), the performance of the method is expected to improve {\color{black} provided there is lesser overlap in the locations of the missing voxels across samples. On the other hand, voxels that are missing across all samples are considered redundant and it is not of interest to estimate the corresponding coefficients.} Such voxels, which may {\color{black}lie in common lesion areas or even outside the brain mask, or exhibit} limited variability across samples, are excluded from {\color{black} further analysis.}

The above features of the proposed tensor response regression model result in distinct advantages over a voxel-wise regression approach, in terms of handling missing voxels. Voxel-wise methods proceed by fitting the model separately corresponding to each voxel, after eliminating the subset of samples for which the corresponding voxel information is missing. Hence the estimation accuracy for the voxel-level coefficients corresponding to missing voxels may deteriorate due to loss of the effective sample size. Further, different subsets of missing voxels across different individuals may potentially result in an imbalance in the effective sample size across voxels. The voxel-wise analysis is expected to be deeply affected by such a reduction in sample size, whereas the tensor-based method is more robust to such issues given the fact that it pools information across voxels to learn tensor margins.

\subsection{Feature Selection} 
The Bayesian approach provides a natural inferential framework that can be used to determine significant effects via credible intervals derived from Markov chain Monte Carlo (MCMC) samples. However, simply computing the $100(1-\alpha)\%$ credible intervals for the parameters of interest in order to determine significance may not provide adequate control for multiplicity and it does not account for the underlying correlations between voxels. Multiplicity adjustments are required when testing for significant effects on a large number of voxels. Typically used adjustments in the Bayesian setting such as the Bonferroni correction  adjusts the significance level with respect to the number of tests (i.e. number of non-missing voxels), but have important limitations. First, the number of MCMC iterations needs to sufficiently exceed the number of voxels for one to construct suitable {\color{black} Bonferroni} corrections. This is typically challenging in the presence of a large number of voxels. Second, the underlying spatial correlations across voxels are not accounted for, which is undesirable for neuroimaging studies. Although alternate approaches such as the CEI method \citep{chumbley2010topological} use post-hoc adjustments to account for spatial correlations at the cluster level resulting in improvements, the performance may still be sensitive to the quality of the estimated coefficients that may be sub-optimal under a voxel-wise regression analysis.

To address these pitfalls, we use joint credible regions for inference and feature selection, which respects the correlations in the posterior distribution and incorporates a naturally in-built multiplicity adjustment mechanism. In particular, we use the `Mdev' method relying on simultaneous credible bands that were introduced by  \cite{crainiceanu2007spatially} and later adopted by \cite{Hua2015}. The joint credible regions are constructed using the posterior samples of the tensor coefficients and respect the shape of the joint posterior distribution and dependence across the coefficients. More concretely, given $J$ posterior samples across $L$ elements of a tensor-valued coefficient, denoted $\{\Gamma^j=(\Gamma(v_1)^j, \ldots,\Gamma^j(v_L))': j=1,\ldots,J\}$ after burn-in, the Mdev method first computes the posterior sample average curve $\widehat{\Gamma(v_l)}$, and pointwise $\alpha/2$ and $1-\alpha/2$ quantiles, $\zeta_{\alpha/2}(v_l)$ and $\zeta_{1-\alpha/2}(v_l)$, respectively, for all $l=1,\ldots,L$. In order to borrow information across voxels jointly and provide in-built multiplicity adjustments, the next step of Mdev involves computing maximal deviations $\zeta^*_{\alpha/2}=\max_{l=1,\ldots,L} \left(\widehat{\Gamma_k(v_l)}-\zeta_{\alpha/2}(v_l) \right)$ and $\zeta^*_{1-\alpha/2}=\max_{l=1,\ldots,L} \left(\zeta_{1-\alpha/2}(v_l)-\widehat{\Gamma_k(v_l)} \right)$. Based on these, the credible bands for each voxel are computed as  $\widehat{\Gamma(v_l)}$ as $\left[\widehat{\Gamma(v_l)}- \zeta^*_{\alpha/2}, \widehat{\Gamma(v_l)}+ \zeta^*_{1-\alpha/2} \right]$. The credible regions naturally provide an inferential framework. For example, the coefficient corresponding to voxel $v_l$ is  considered significant if the credible interval $\left[\widehat{\Gamma(v_l)}- \zeta^*_{\alpha/2}, \widehat{\Gamma(v_l)}+ \zeta^*_{1-\alpha/2} \right]$ does not contain zero; otherwise it is considered to be {\color{black} non-significant}. 
Extensive numerical experiments show superior control over false positives (higher specificity) compared to analysis without multiplicity adjustments \citep*{eklund2016cluster}, and superior power to detect important features compared to voxel-wise methods with Bonferroni corrections, under the proposed joint credible regions approach. As a byproduct, the credible intervals can also be used for uncertainty quantification, which is another desirable quality in neuroimaging studies.  

\subsection{Inferring Group and Individual Neuro-plasticity Brain Maps}

Using the BTRR approach, it is possible to infer both group-level as well as individual neuroplasticity maps indicating voxels with significant neuroplasticity. The latter clearly provides deeper insights compared to a standard group-level comparison that is typically reported in neuroplasticity studies, and it is of independent interest for studying personalized trajectories of 
{\color{black}response to treatment}. 


Under the BTRR model (\ref{eq:vxl-longBTRR}), the individual-level neuroplasticity is derived from MCMC samples, and these quantities are subsequently used to compute group-level neuroplasticity. For the $v$th voxel and $j$th MCMC iteration, we compute neuroplasticity between time points $t$ and $t^*=t-1$ as:
\begin{align}
    & \hat{\Delta}^{(j)}_{t,t^*,i}(v)= \hat{y}^{(j)}_{ti}(v)-\hat{y}^{(j)}_{t^*i}(v) \nonumber \\ 
    & = \left(\widehat{\Gamma}^{(j)}(v)+\widehat{\Theta}^{(j)}_i(v) \right) \left(\mathcal{T}_{ti}-\mathcal{T}_{t^*i} \right) + \sum_{m=1}^M \left(\widehat{\mathcal{B}}_{tm}^{(j)}-\widehat{\mathcal{B}}_{t^*m}^{(j)} \right) c_{im} + {\color{black} \sum_{q=1}^Q \widehat{\mathcal{C}}_q^{(j)} (z_{tiq}-z_{t^*iq})}
    \label{eq:ind-plasticity}
\end{align}

\noindent for $i=1,\ldots,n,$ where $\widehat{\Gamma}^{(j)},\widehat{\Theta}^{(j)}_i,\widehat{\mathcal{B}}^{(j)}_{tm}$, {\color{black} and $\widehat{\mathcal{C}}^{(j)}_q$} represent the values of $\Gamma, \Theta_i, \mathcal{B}_{tm}$, {\color{black} and $\mathcal{C}_q$} at the $j$th iteration, and the differences in (\ref{eq:ind-plasticity}) are stored over all MCMC iterations $j=1,\ldots,J$. Subsequently, the joint credible regions approach in Section 2.3 is used to infer the voxels with significant neuroplasticity changes, based on the differences in (\ref{eq:ind-plasticity}) over all MCMC iterations after burn-in. 

In order to obtain the group-level neuroplasticity maps, we average over the individual neuroplasticity maps within a given group. Consider two groups $\mathcal{G}_0, \mathcal{G}_1$ of interest. Then the neuroplasticity maps corresponding to $\mathcal{G}_0$ is obtained using the MCMC samples $\big\{\hat{\Delta}^{(j)}_{t,t^*,\mathcal{G}_1}(v)=\sum_{i \in \mathcal{G}_1}\hat{\Delta}^{(j)}_{t,t^*,i}(v)/n_{\mathcal{G}_1} : j=1,\ldots,J \big\}$, and similarly for the other group $\mathcal{G}_1$. For group $\mathcal{G}$, the MCMC samples $\big\{\hat{\Delta}^{(j)}_{t,t^*,\mathcal{G}}(v),j=1,\ldots,J,v=1,\ldots,V\}$ can be used to determine significant neuroplasticity maps based on the joint credible regions method in Section 2.3. We note that the groups under consideration are pre-determined based on the scientific questions of interest for a given application. 

\section{Posterior Computation}
The model parameters in  (\ref{eq:longBTRR})-(\ref{eq:longBTRRprior}) are unknown and estimated via posterior distributions that are obtained by combining the prior and the likelihood under the Bayesian paradigm. Since it is typically challenging to obtain closed-form joint posterior distributions for the model parameters, a Markov chain Monte Carlo (MCMC) sampling scheme is employed that is able to draw samples from the joint posterior in a computationally cohesive manner. Under the MCMC scheme, it is possible to estimate the full posterior distribution that is then used to derive point estimates and provide uncertainty quantification. The MCMC steps for most parameters in the model such as the tensor margins and the scale parameters involve fully Gibbs updates with closed-form posterior distributions, that result in good mixing. The only parameters that can not be updated using fully Gibbs steps involve the lengthscale parameters ($\alpha$) in the covariance matrices $\mathbf{W}$ for the tensor margins, which are updated using Metropolis-Hastings steps. Under the assumed prior structure, the number of such lengthscale parameters is reasonable, and a Metropolis random walk is used for updating these parameters. In particular, we use the proposal density $\alpha_{d,r,s_x+1}^{\mu} | \alpha_{d,r,s_x}^{\mu} \sim \text{log-Normal}(\alpha_{d,r,s_x}^{\mu}, \sigma^2_{\alpha})$, where $s_x$ indexes the MCMC iteration, and $\sigma^2_{\alpha}$ denotes the variance term that is fixed. We run the Gibbs sampling steps for a total of 5000 MCMC iterations with a burn-in of 2500. 

\vskip 10pt

{\noindent \underline{Choosing the tensor rank (R):}} We use the deviance information criterion (DIC) to select the tensor rank $R$, which is a goodness of fit measure that strikes a balance between the quality of the model fit and the number of parameters involved in the model \citep{shriner2009deviance}. Such a criterion was also used in \cite{Guhaniyogi2021}, and resulted in suitable performance in a wide variety of studies. 

\section{Simulation Studies}

\subsection{Data Generation Schemes}

Model performance was conducted using several distinct data generation schemes {\color{black} (see Table \ref{tab:Table2})}. For each scheme, {\color{black} 50 replicates} of simulated longitudinal data were generated according to Model (\ref{eq:longBTRR}) but without a subject-specific time slope (i.e. $\Theta_i=0$). The covariates $x_{is}$ {\color{black} and $z_{tiq}$} and noise terms $\mathcal{E}_{ti}$ were simulated from Gaussian distributions. For each scheme, tensor-valued outcomes $\mathcal{Y}_{ti}$ of size $16 \times 16 \times 16$ were generated (total of 4096 voxels) across 14 subjects (indexed by $i=1,\ldots,14$), and for 3 longitudinal visits (indexed by $t=0,1,2$). 
We specified the mean signal-to-noise (SNR) ratio across all voxels as approximately $0.75$, 
where a lower SNR signals greater noise in the images.  Additional SNR levels were also investigated and produced similar results, but they are not presented here due to space constraints. After generation, the data was split into a training set that contained all voxels from the first two visits and a randomly selected subset of 75\% and 50\% voxels in the third visit, and a test set that contained the remaining holdout voxels (25\% and 50\%) in the third visit.  The proposed model was fit on the training data and feature selection results were reported based on this fit, whereas the trained model was used to make out of sample prediction on the test set data. The reported results were averaged over replicates.

 The data generation schemes differed in terms of the structure of the tensor-valued coefficients in model (\ref{eq:longBTRR}). In Scheme 1, no time-varying coefficients were assumed {\color{black}(i.e. $\mathcal{B}_{tm} = 0$)}, and {\color{black} all other coefficients (including the intercept terms) were generated from rank 2 tensor decompositions with Binomially-distributed tensor margins for 4 different covariates}. For each tensor margin, the probability that a given element would be zero was fixed at {\color{black} $0.55$} such that after constructing the low-rank tensor coefficients, approximately 75\% of voxels in each coefficient consisted of true zeros across replicates. Scheme 2 was similar to Scheme 1 in terms of not having any time-varying effects, but differed with regard to how the remaining coefficients were generated. In particular, these coefficients were set to be non-zero for approximately-spherical (Scheme 2a) and {\color{black} cubic} (Scheme 2b) volumes of voxels with randomly chosen centers. The volumes of these shapes were fixed such that an average of {\color{black} 75\%} of voxels consisted of true zeros across coefficients. 
 
 In contrast to Schemes 1 and 2, data under the Scheme 3  was generated based on both time-varying as well as time-invariant effects. {\color{black} Three covariates} were chosen to have time-invariant effects in Scheme 3, and the pattern of these effects was chosen similarly as in Schemes 2a and 2b, corresponding to Schemes 3a and 3b respectively. Only one covariate (generated from a binomial distribution) was assigned to exhibit time-varying effects in Scheme 3, and these effects assumed spherical (Scheme 3a) and cubic (Scheme 3b) shapes. The volume of these time-varying shapes {\color{black} varied} across the three visits, with the proportion of true zeros increasing from 50\% at the first visit to 93\% at the third visit. 
Examples of the different classes of signals in Schemes 1-3 are summarized in Table \ref{tab:Table2}, and visually illustrated in Figures \ref{fig:Figure2}-\ref{fig:Figure3}.

\begin{table}[H]
    \centering
    \begin{tabular}{*6c}
        \toprule
        Setup Options & Scheme 1 & Scheme 2A & Scheme 2B & Scheme 3A & Scheme 3B \\
        \midrule
        Coefficient Type & Low-rank & Spherical & Cubic & Spherical & Cubic \\
        Number of time-varying coefficients & 0 & 0 & 0 & 1 & 1 \\
        Total number of covariates & 4 & 4 & 4 & 4 & 4 \\
        Number of time-varying covariates & 2 & 2 & 2 & 2 & 2 \\
        Percentage of true coefficient zeros & 75\% & 75\% & 75\% & 50-93\% & 50-93\% \\
        Mean SNR & $0.75$ & $0.75$ & $0.75$ & $0.75$ & $0.75$ \\
        \bottomrule
    \end{tabular}
    \caption{Summary of the various simulation schemes used.}
    \label{tab:Table2}
\end{table}
 
\begin{figure}[!t]
\centering
\begin{subfigure}[t]{.3\textwidth}
    \centering
    \includegraphics[width=\linewidth]{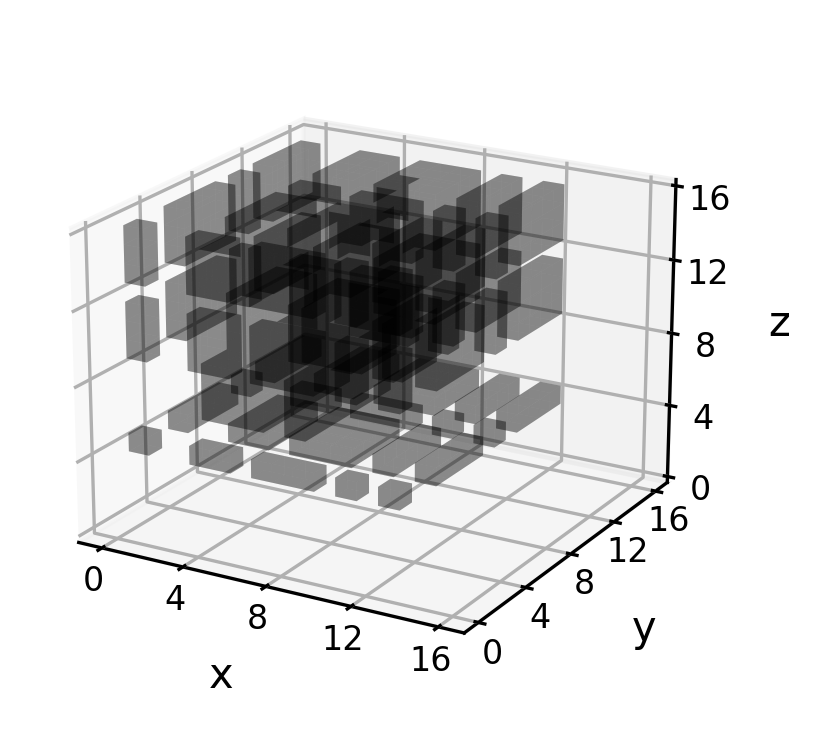}
    \caption{}
  \end{subfigure}
  \hfill
  \begin{subfigure}[t]{.3\textwidth}
    \centering
    \includegraphics[width=\linewidth]{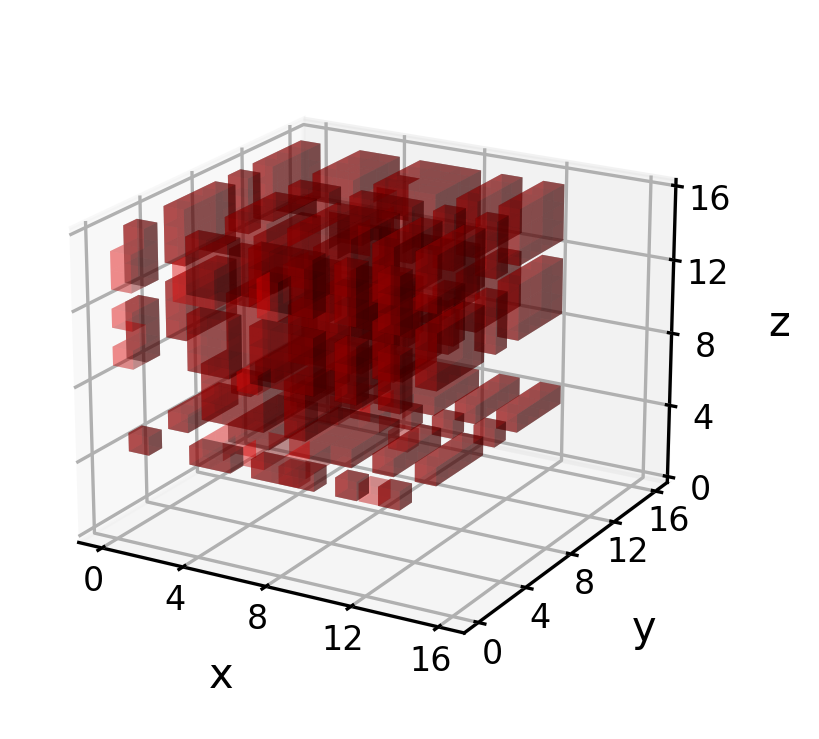}
    \caption{}
  \end{subfigure}
  \hfill
  \begin{subfigure}[t]{.3\textwidth}
    \centering
    \includegraphics[width=\linewidth]{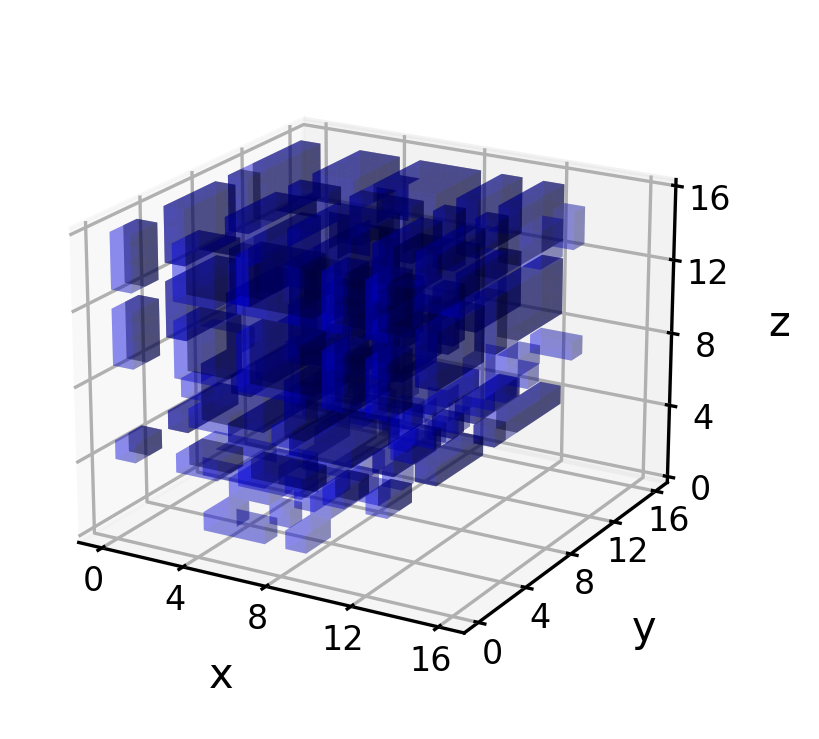}
    \caption{}
  \end{subfigure}
  \begin{subfigure}[t]{.3\textwidth}
    \centering
    \includegraphics[width=\linewidth]{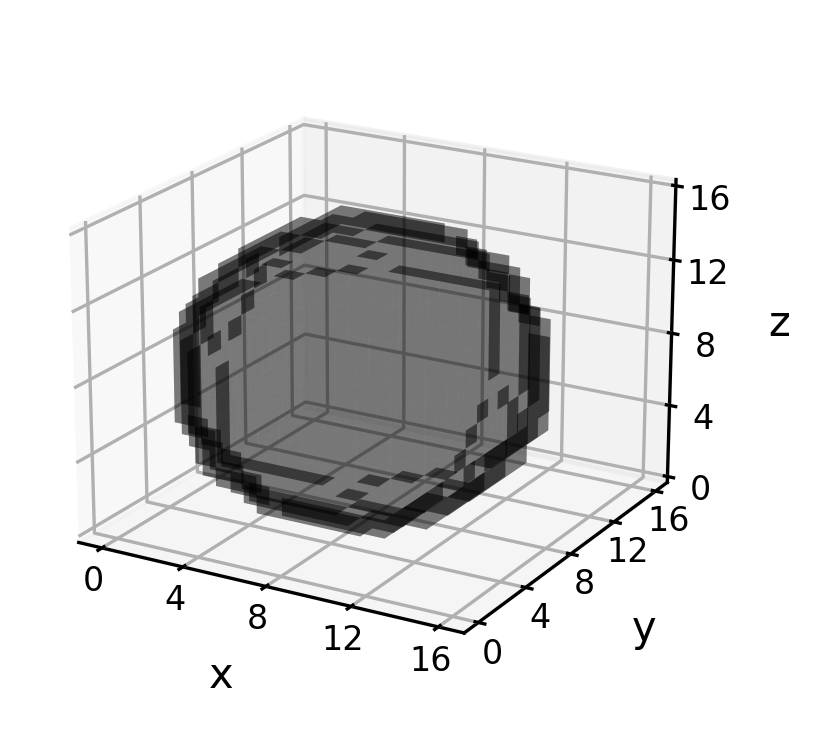}
    \caption{}
  \end{subfigure}
  \hfill
  \begin{subfigure}[t]{.3\textwidth}
    \centering
    \includegraphics[width=\linewidth]{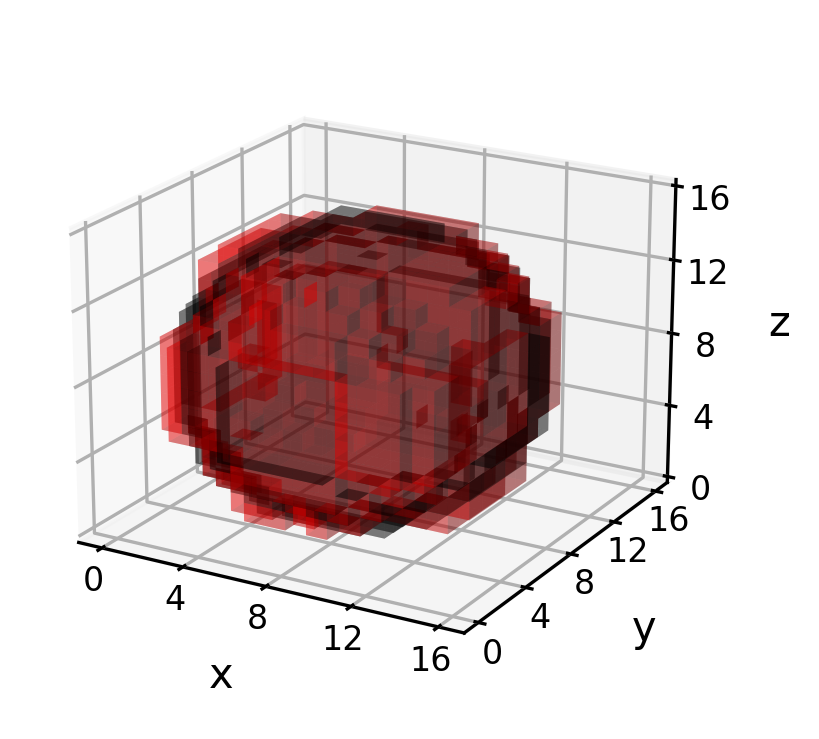}
    \caption{}
  \end{subfigure}
  \hfill
  \begin{subfigure}[t]{.3\textwidth}
    \centering
    \includegraphics[width=\linewidth]{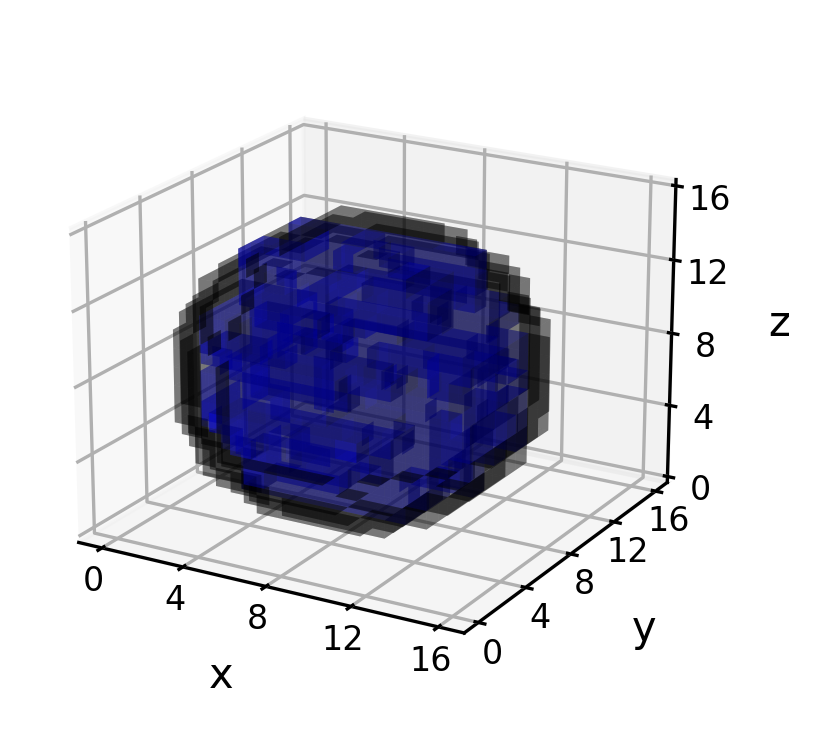}
    \caption{}
  \end{subfigure}
  \medskip
    \begin{subfigure}[t]{.3\textwidth}
    \centering
    \includegraphics[width=\linewidth]{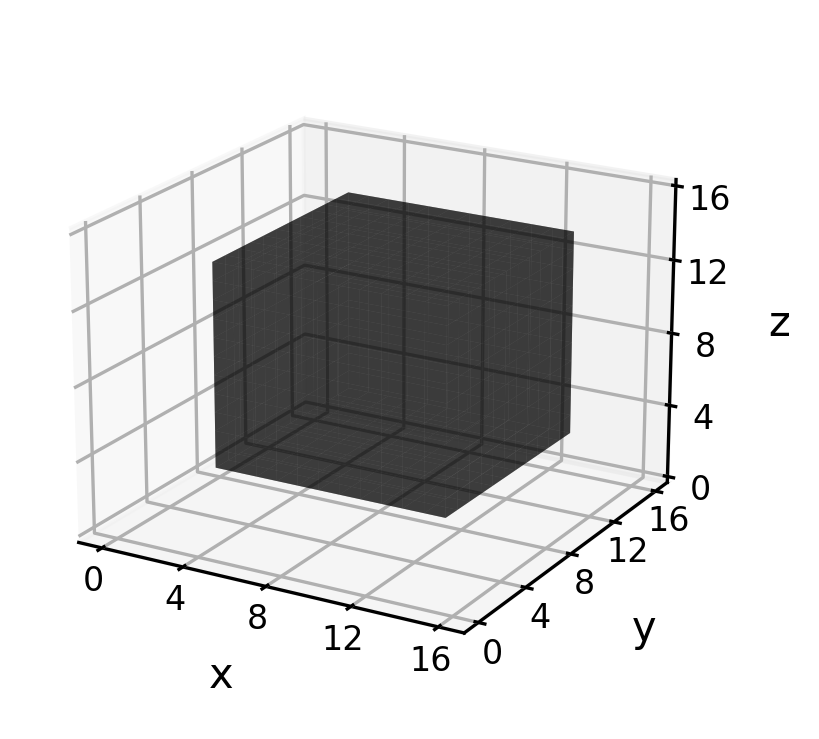}
    \caption{}
  \end{subfigure}
  \hfill
  \begin{subfigure}[t]{.3\textwidth}
    \centering
    \includegraphics[width=\linewidth]{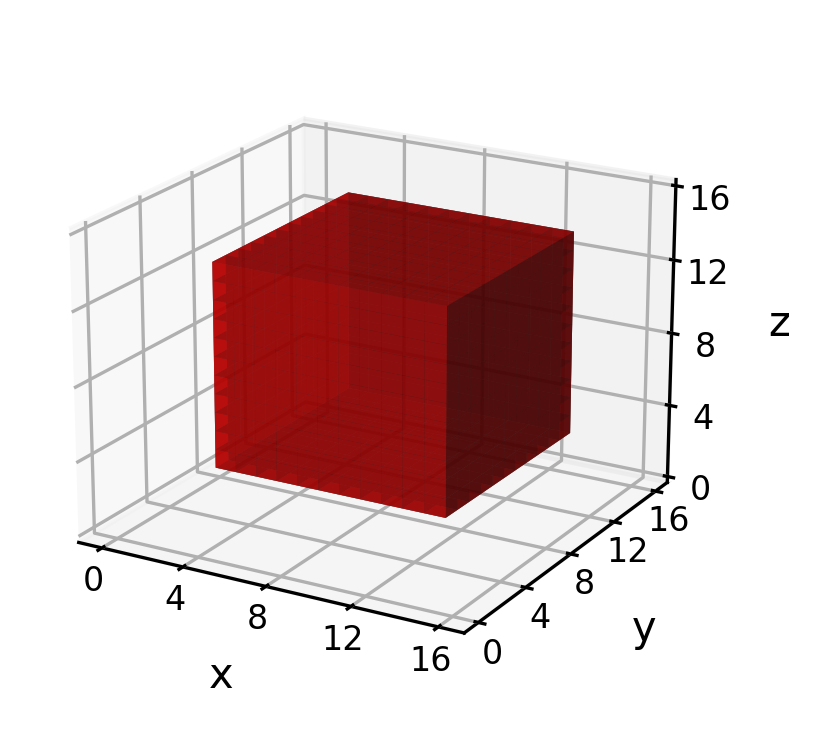}
    \caption{}
  \end{subfigure}
  \hfill
  \begin{subfigure}[t]{.3\textwidth}
    \centering
    \includegraphics[width=\linewidth]{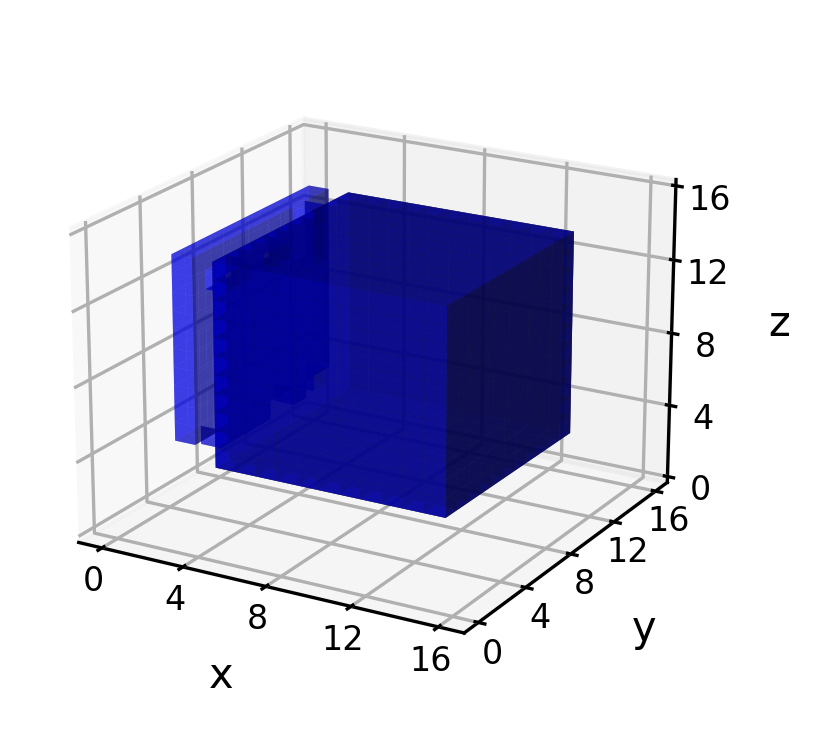}
    \caption{}
  \end{subfigure}
  \caption{\textbf{Scheme 1-2 signals}. Panels (a), (d), \& (g) -- True coefficients for a chosen Scheme 1, Scheme 2a, and Scheme 2b replicate, respectively (all time-invariant); Panels (b), (e), \& (h) -- Estimated coefficients using l-BTRR method (red) overlaid on true signal (black); Panels (c), (f), \& (i) -- Estimated coefficients using cs-BTRR method (blue) overlaid on true signal (black).}
  \label{fig:Figure2}
\end{figure}

\begin{figure}[!t]
  \begin{subfigure}[t]{.16\textwidth}
    \centering
    \includegraphics[width=\linewidth]{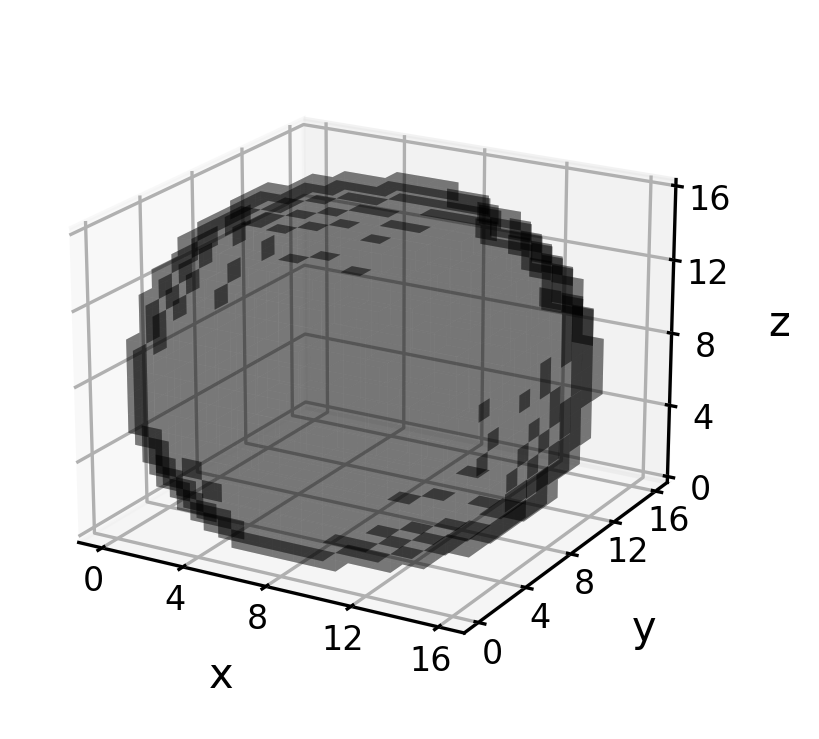}
  \end{subfigure}
  \begin{subfigure}[t]{.16\textwidth}
    \centering
    \includegraphics[width=\linewidth]{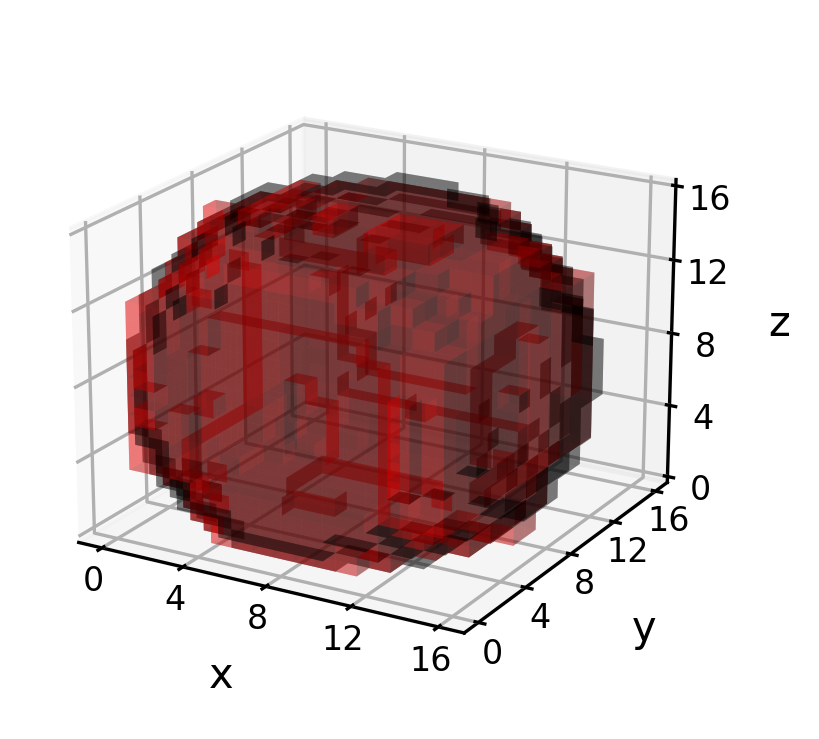}
  \end{subfigure}
  \begin{subfigure}[t]{.16\textwidth}
    \centering
    \includegraphics[width=\linewidth]{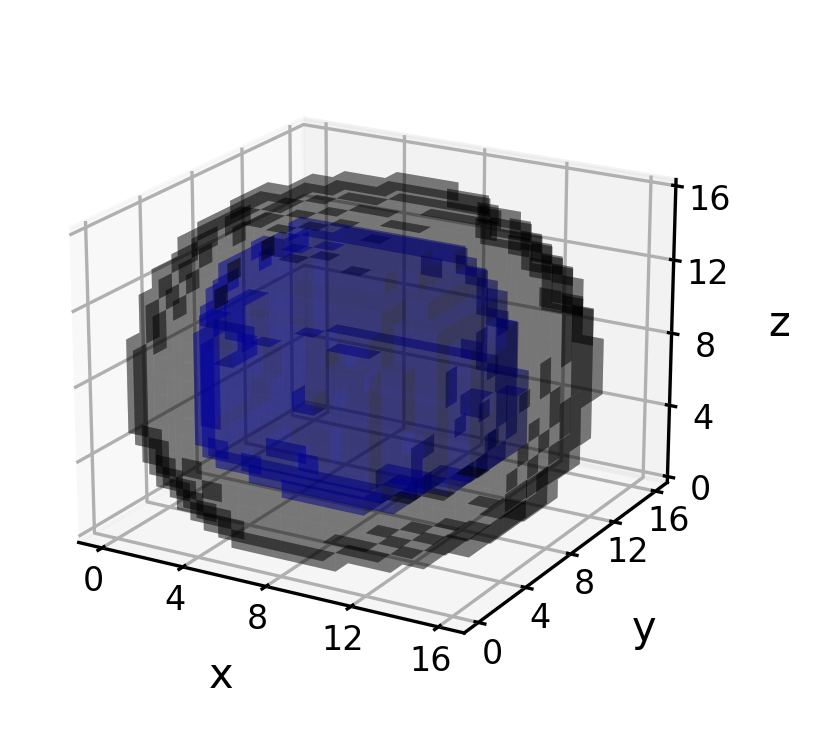}
  \end{subfigure}
    \begin{subfigure}[t]{.16\textwidth}
    \centering
    \includegraphics[width=\linewidth]{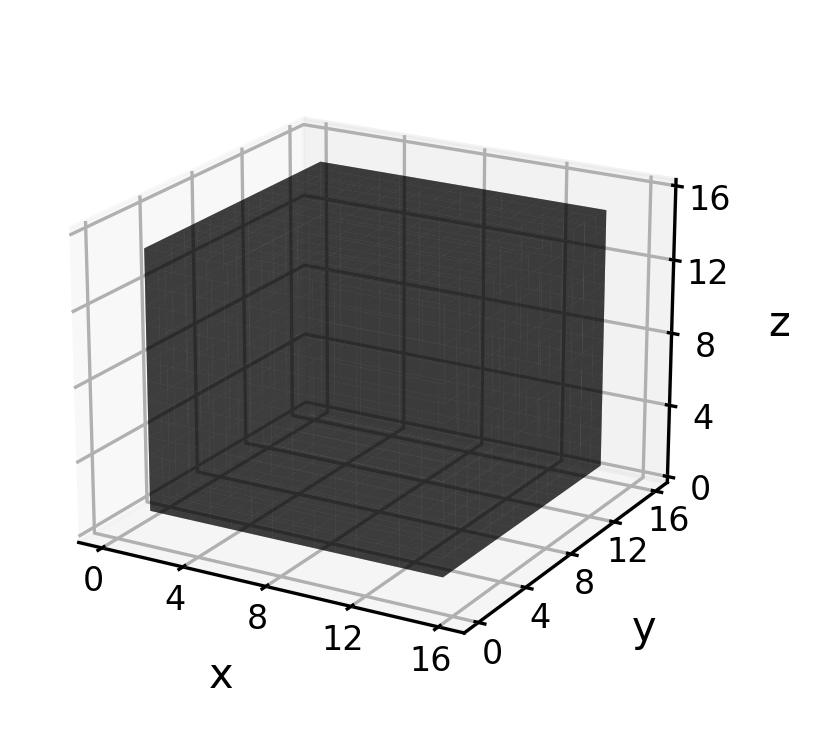}
  \end{subfigure}
  \begin{subfigure}[t]{.16\textwidth}
    \centering
    \includegraphics[width=\linewidth]{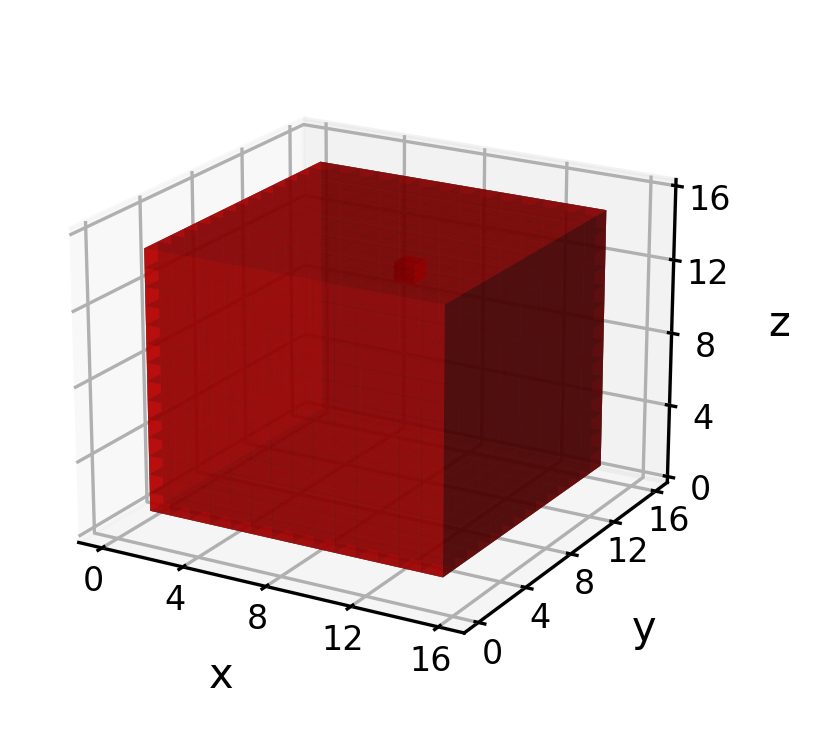}
  \end{subfigure}
  \begin{subfigure}[t]{.16\textwidth}
    \centering
    \includegraphics[width=\linewidth]{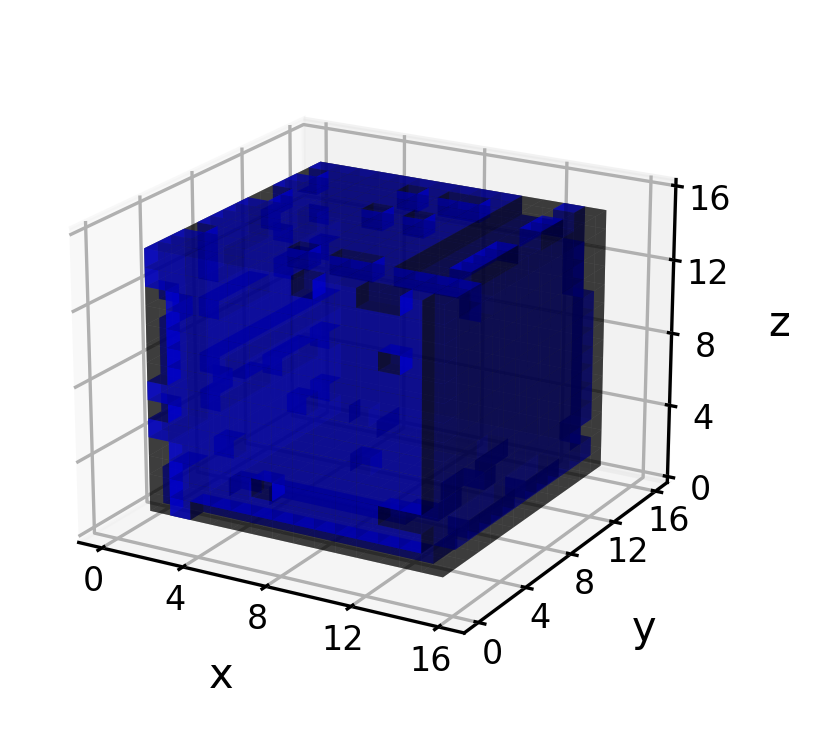}
  \end{subfigure}
  \medskip
    \begin{subfigure}[t]{.16\textwidth}
    \centering
    \includegraphics[width=\linewidth]{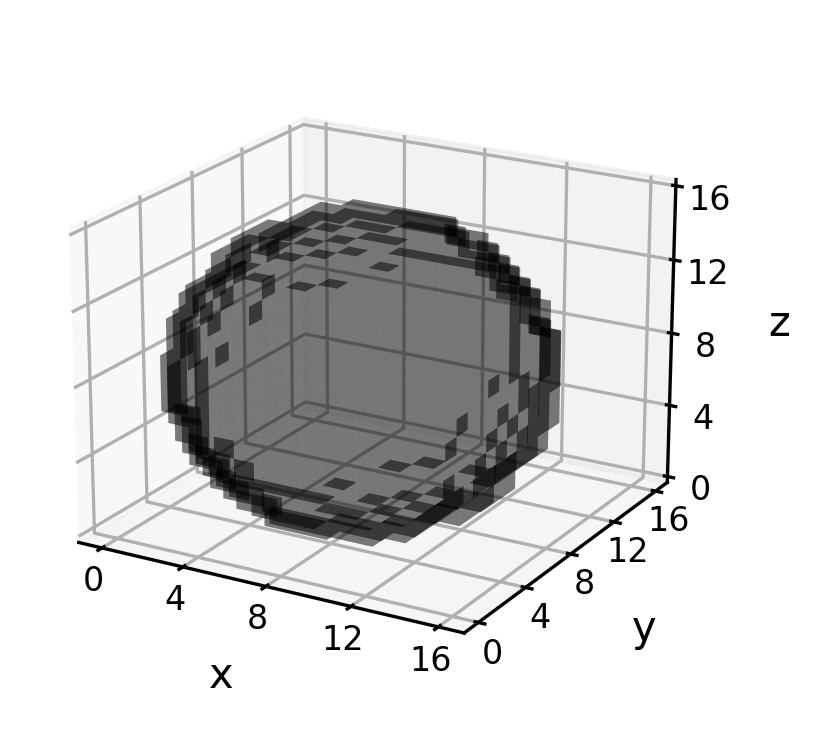}
  \end{subfigure}
  \begin{subfigure}[t]{.16\textwidth}
    \centering
    \includegraphics[width=\linewidth]{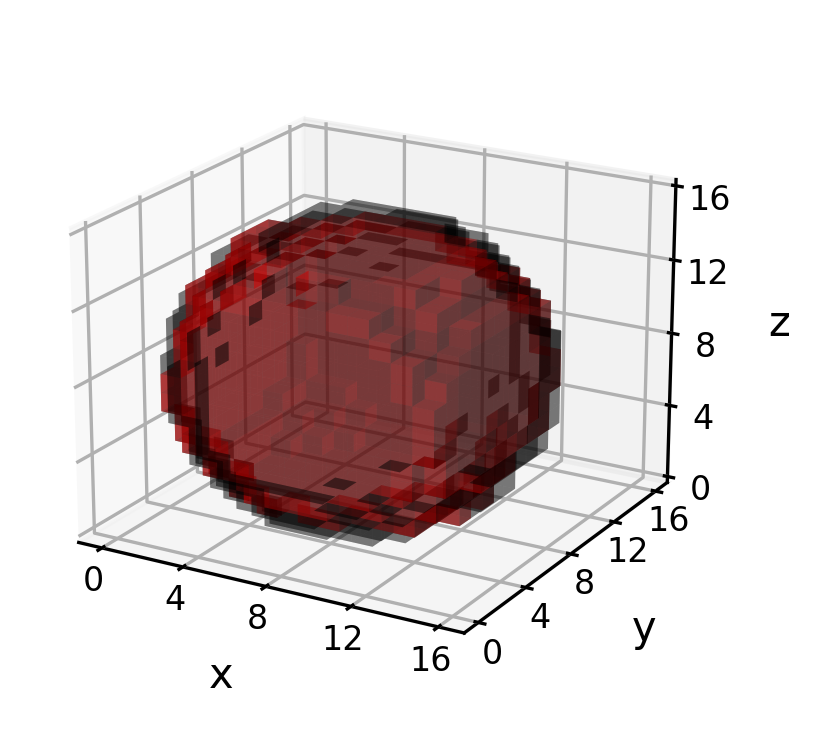}
  \end{subfigure}
  \begin{subfigure}[t]{.16\textwidth}
    \centering
    \includegraphics[width=\linewidth]{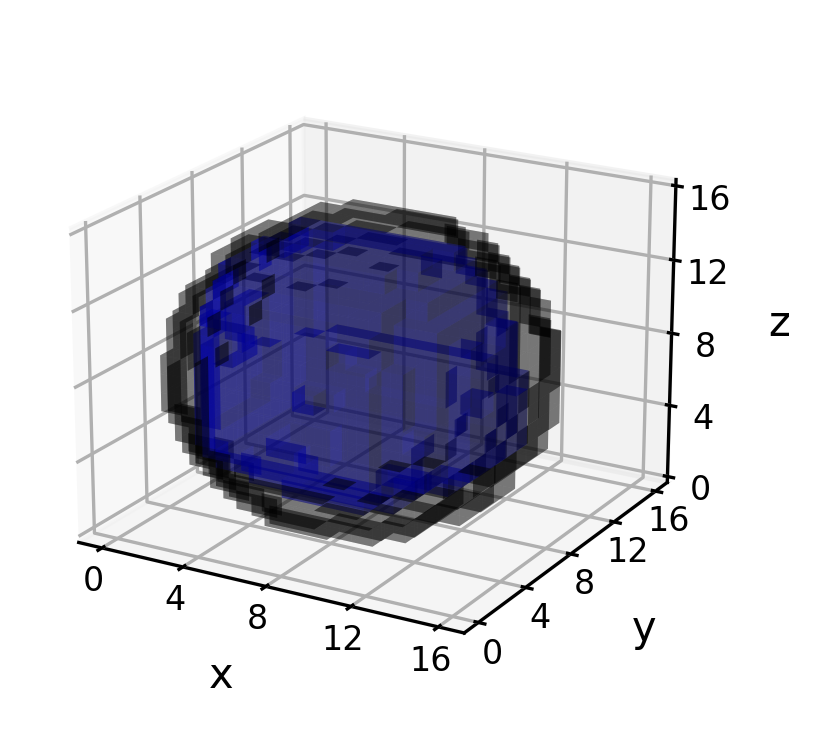}
  \end{subfigure}
  \begin{subfigure}[t]{.16\textwidth}
    \centering
    \includegraphics[width=\linewidth]{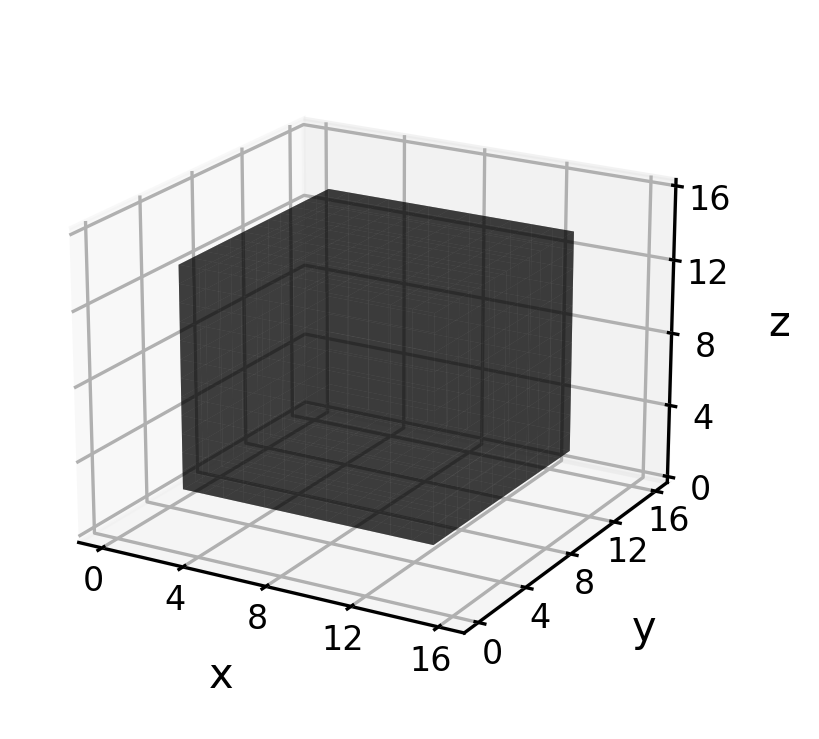}
  \end{subfigure}
  \begin{subfigure}[t]{.16\textwidth}
    \centering
    \includegraphics[width=\linewidth]{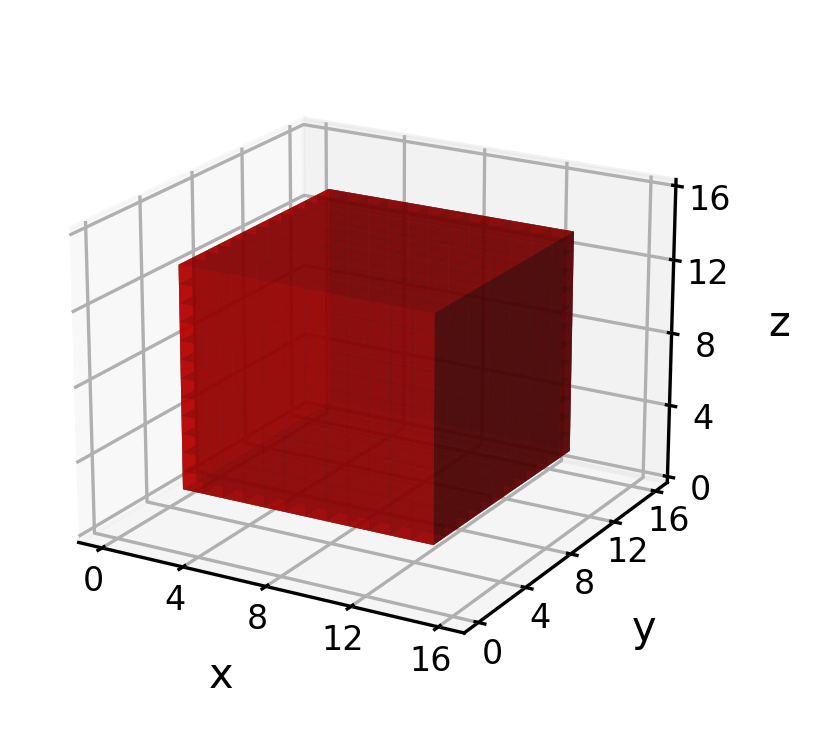}
  \end{subfigure}
  \begin{subfigure}[t]{.16\textwidth}
    \centering
    \includegraphics[width=\linewidth]{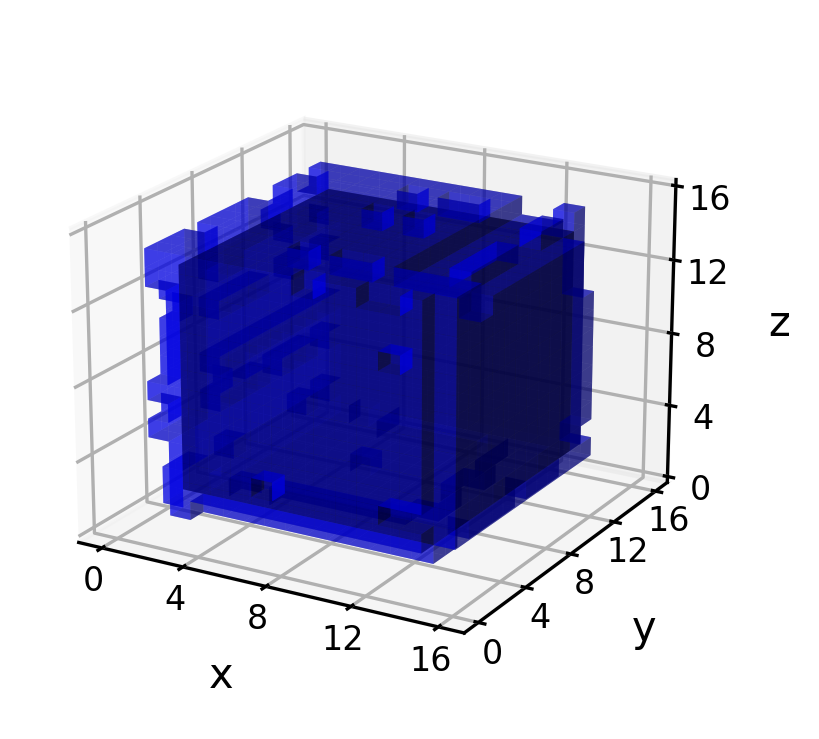}
  \end{subfigure}
  \medskip
    \begin{subfigure}[t]{.16\textwidth}
    \centering
    \includegraphics[width=\linewidth]{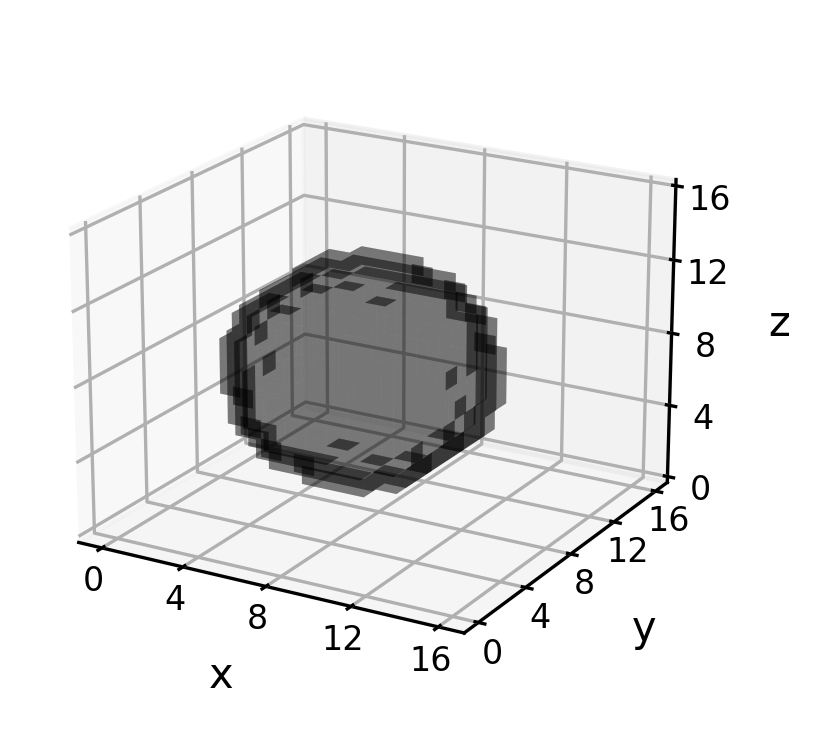}
  \end{subfigure}
  \begin{subfigure}[t]{.16\textwidth}
    \centering
    \includegraphics[width=\linewidth]{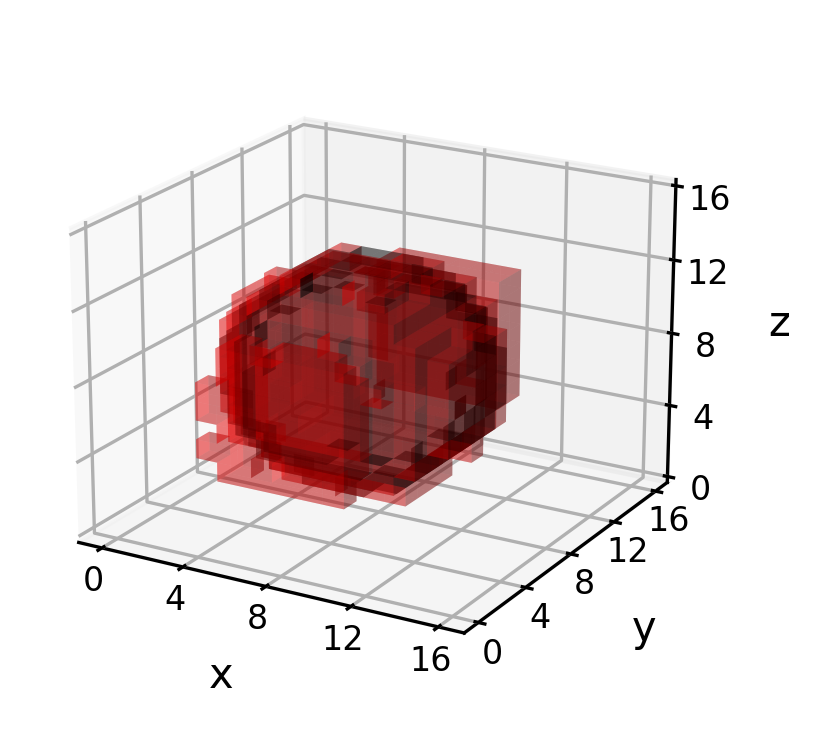}
    \caption{}
  \end{subfigure}
  \begin{subfigure}[t]{.16\textwidth}
    \centering
    \includegraphics[width=\linewidth]{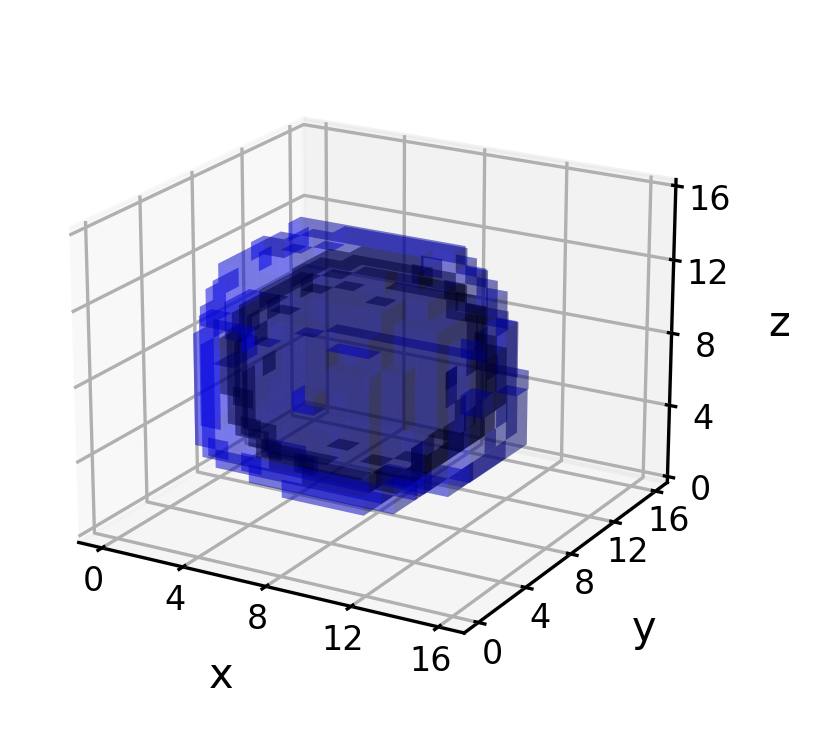}
  \end{subfigure}
    \begin{subfigure}[t]{.16\textwidth}
    \centering
    \includegraphics[width=\linewidth]{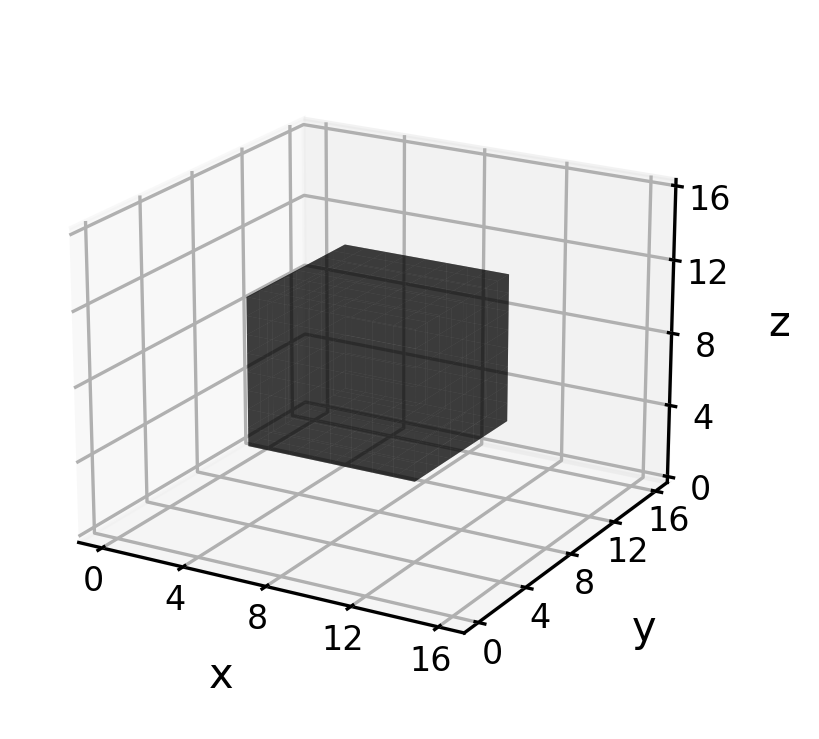}
  \end{subfigure}
  \begin{subfigure}[t]{.16\textwidth}
    \centering
    \includegraphics[width=\linewidth]{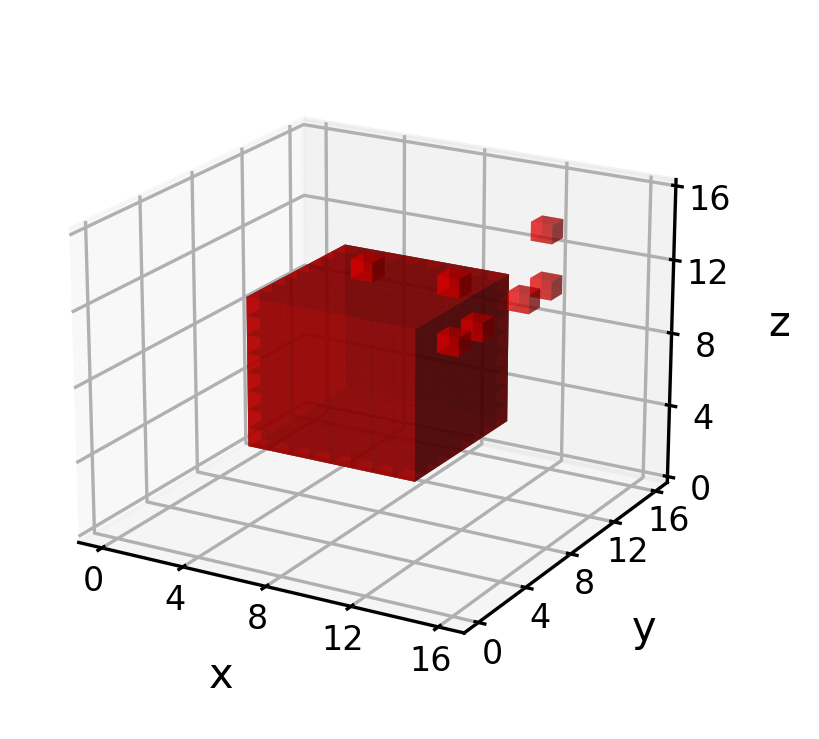}
    \caption{}
  \end{subfigure}
  \begin{subfigure}[t]{.16\textwidth}
    \centering
    \includegraphics[width=\linewidth]{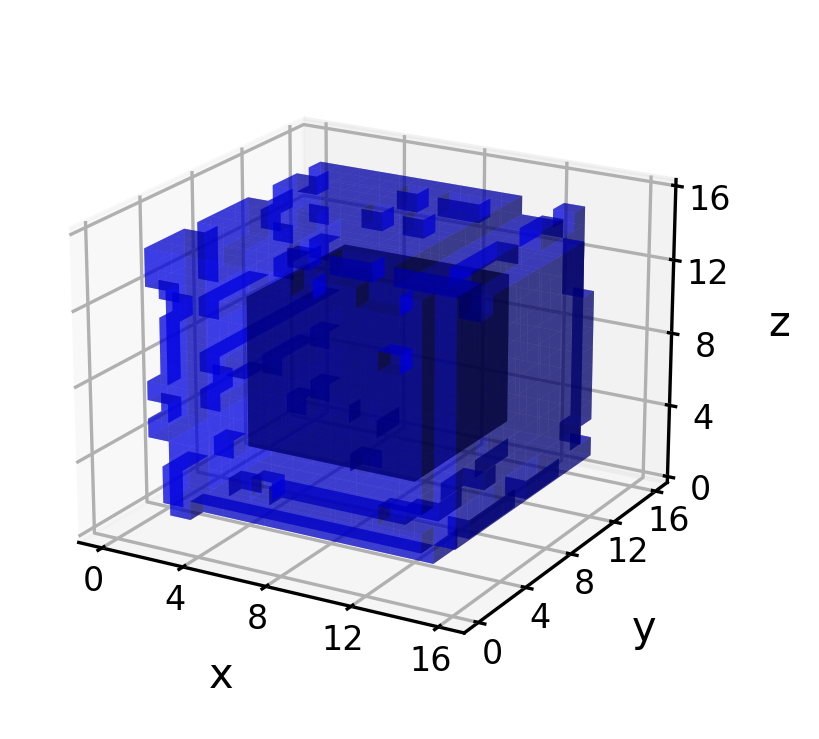}
  \end{subfigure}
  \caption{\label{fig:Figure3} {\textbf{Scheme 3 signals}. Panel (a). True and estimated time-varying signals for a chosen Scheme 3a replicate; rows 1-3 pertain to longitudinal visit (0-2), and columns 1-3 pertain to true signals (black), signals estimated with l-BTRR (red), and signals estimated with cs-BTRR (blue) respectively.  Panel (b). True and estimated time-varying signals for a chosen Scheme 3b replicate; rows 1-3 pertain to longitudinal visit (0-2), and columns 1-3 pertain to true signals (black), signals estimated with l-BTRR (red), and signals estimated with cs-BTRR (blue) respectively.}}
\end{figure}

\subsection{Competing Methods and Performance Metrics}

To compare the performance of the longitudinal BTRR (l-BTRR) method, five competing methods were used that can be categorized into cross-sectional and longitudinal methods, {\color{black} as well as tensor-based modeling and voxel-wise approaches that are routinely used}. The only competing approach that accounted for the spatial configuration of the voxels was the cross-sectional BTRR (cs-BTRR), which fits the Bayesian tensor response regression model by ignoring the within-subject dependence across longitudinal visits, i.e. treating the longitudinal visits as exchangeable. Hence this approach does not pool information across visits to estimate subject-level effects. The cs-BTRR approach is obtained by modifying model (\ref{eq:longBTRR}) to exclude the effects corresponding to $B_i$, $\Gamma$, and $\Theta_i$, and further assuming $\mathcal{B}_{tm}$ is the same for all visits $t$, which translates to a  time-invariant effect for all covariates. 
Both the l-BTRR and cs-BTRR methods consisted of 5000 MCMC iterations per replicate, using a burn-in of 2500. {\color{black} For each replicate, tensor coefficient ranks between 1-5 were considered, where each coefficient rank was set as equal for simplicity. Optimal rank was determined by finding the minimal DIC score. To yield more comparable evaluation metrics between the l-BTRR and cs-BTRR methods, we performed the DIC rank selection for the l-BTRR method, and that chosen rank was manually specified for the cs-BTRR approach.} For the Bayesian methods, significant features were inferred by computing the joint simultaneous credible bands as per Section 2.3, where Type I error rate was set to $\alpha=0.05$. 

The remaining four competing methods are all voxel-wise approaches, meaning they perform the analysis separately for each voxel. {\color{black}The voxel-wise approaches were fit using either the ordinary least squares (OLS) technique, or the Lasso approach \citep{tibshirani1996regression}.} These methods include cross-sectional approaches such as voxel-wise cross-sectional ordinary least squares (vcs-OLS) and voxel-wise cross-sectional Lasso (vcs-Lasso), as well as longitudinal methods such as the voxel-wise linear mixed modeling (vl-OLS), and the voxel-wise longitudinal Lasso with random intercepts (vl-Lasso). Unlike l-BTRR, the longitudinal voxel-wise approaches (i.e. vl-OLS and vl-Lasso) fit the voxel-level model (\ref{eq:vxl-longBTRR}) separately for each voxel, 
{\color{black} while still incorporating subject and time-dependence through a random intercept and time-varying effects as in (\ref{eq:vxl-longBTRR})}.
This construction is consistent with linear mixed modeling for longitudinal data literature \citep{curran2011disaggregation}. On the other hand, the voxel-wise cross-sectional methods ignored the within-subject dependence across visits and assume time-invariant effects across all visits. The voxel-wise Lasso approaches were fit using the {\it glmmLasso} and {\it glmnet} R packages, and resulted in sparse estimates for the regression coefficients. For both vl-Lasso and vsc-Lasso, the tuning parameter was selected as the parameter which produced minimum cross-validation error in a 10-fold cross-validation scheme. 



For each Scheme and each chosen level of holdout voxels ($25\%$ and $50\%$), performance metrics were selected to assess the out-of-sample predictive performance and feature selection accuracy. For out-of-sample prediction, overall root mean squared error (p-RMSE) was calculated by averaging the squared error across all subjects, visits, and hold-out voxels. Additionally, out-of-sample correlation (p-Corr) was computed comparing the vectorized observed and predicted outcome for all subjects, visits, and hold-out voxels. Probability of coverage was computed for each method by obtaining multiplicity-adjusted credible or confidence intervals on the fitted outcome and determining the proportion of holdout voxels whose interval contained the true value (without simulated noise) across replicates. The average width of these confidence and credible intervals are presented along with coverage probability to get a sense of the precision of each method.} {\color{black} Coefficient estimation accuracy was obtained by computing the RMSE between the true and estimated coefficients across voxels (c-RMSE).} For feature selection, sensitivity, specificity, and F$1$ score were calculated by comparing significance estimates  with true coefficient values. In particular, the F$1$ score is computed as the harmonic mean between recall (sensitivity) and precision, i.e. 2 recall$\times$precision/(recall+precision). Here, sensitivity is the power to detect the true positives, computed as the proportion of truly non-zero coefficients that were inferred as significant, and precision is defined as the ratio of the number of true positives over the total number of significant signals detected. We also report specificity which is defined as the proportion of truly zero coefficients that were correctly identified as such. {\color{black} All feature selection metrics were averaged for the coefficients $\mathcal{B}_{tm}$, $\mathcal{D}_s$, and $\mathcal{C}_q$ in Model \ref{eq:longBTRR}, which correspond to population-level effects of simulated covariates $c_{im}$, $x_{is}$, and $z_{tiq}$}.

The significance coefficient estimates are obtained differently for each class of methods. For the BTRR methods, simultaneous credible bands are used to infer significant voxel-specific estimates from MCMC samples (see Section 2.3), and point-wise credible bands without multiplicity correction are computed for comparison. The simultaneous credible bands for the Bayesian tensor approach automatically result in multiplicity adjustments. {\color{black} For OLS methods, we implement a version of cluster-extent inference (CEI) that utilizes the Benjamini-Hochberg procedure to adaptively select a threshold that distinguishes between significant and non-significant clusters of voxels \citep{chumbley2010topological}. {\color{black} In order to compute coverage probabilities for the voxel-wise methods, we first fit the model for each holdout voxel using training data from the first and second visit and subsequently obtained the multiplicity-adjusted confidence intervals for prediction of the outcome using covariates from the third visit.} For Lasso methods, all effects with a magnitude greater than $0.001$ were deemed to be significant for a given voxel. Hence, we were not able to include multiplicity adjustments for the Lasso approaches, {\color{black} and therefore the corresponding probability of coverage was not computed.}

\subsection{Simulation Results}

The simulation results are presented in {\color{black} Tables \ref{tab:Table3}-\ref{tab:Table5}}. {\color{black} The proposed longitudinal BTRR method} has a significantly lower out-of-sample prediction RMSE, significantly higher correlation for the predicted outcome values, and consistently improved regression coefficient estimation {\color{black}(lower c-RMSE values)} compared to all other competing methods and across all settings. These results point to the clear advantages in prediction and parameter estimation under the proposed longitudinal tensor approach. {\color{black} The higher c-RMSE values in the voxel-wise methods correspond to more isolated patterns in the estimated coefficients that fail to preserve spatial contiguity that is seen in the true signal. In contrast, the tensor-based methods are able to preserve the spatially-distributed signals in the estimated coefficients, as seen in Figures \ref{fig:Figure2}-\ref{fig:Figure3}. Additionally, these figures illustrate that the tensor-based methods are adaptive to the spatial discontinuity of the signals, given that the estimated signals preserve the sharp edges of the true signal, especially for the l-BTRR method. To explore this further, we examined the coefficient estimation and feature selection of all competing methods within a thin strip around the perimeter of the true signals in Scheme 2, where the discontinuity is most pronounced. These results are consistent with the overall results and are shown in Table \ref{tab:Table6}.} 

In terms of feature selection, the proposed method consistently has the highest F1 score, which validates the superior performance. {\color{black} The cross-sectional BTRR approach and longitudinal Lasso method often report the highest sensitivity that exceeds the sensitivity under the multiplicity-adjusted l-BTRR method. 
} While this is expected, the higher sensitivity reported under the Lasso methods {\color{black}(without multiplicity correction)} comes at the cost of considerably lower specificity that also percolates to significantly lower F1 score under these approaches. In contrast, the F1 score under the proposed approach is always greater than $0.75$. {\color{black} The proposed method reports the second highest specificity after voxel-wise OLS methods for all schemes.} We note that the specificity values for the l-BTRR method correspond to false discovery rates which are relatively close to the nominal value of $0.05$, which is expected given the built-in multiplicity adjustment under the simultaneous credible bands. The voxel-wise OLS methods always have {\color{black} specificity close to 1} after applying {\color{black} CEI} multiplicity correction, which is due to the fact that it detects very few significant features as evidenced by low sensitivity values. This is clearly not desirable,{\color{black} and indicates the pitfalls of voxel-wise model fitting for feature selection in simulated longitudinal cases with small sample size and relatively large numbers of voxels.} {\color{black}The considerably poor performance under the voxel-wise approaches also stems from their inability to pool information across voxels and to account for their spatial configurations.} Overall, the proposed l-BTRR method illustrates robust and accurate feature selection in the presence of time-varying as well as time-invariant signals {\color{black} and covariates.}


{\color{black}Moreover,} the multiplicity-adjusted cs-BTRR performs the second best in general, consistently registering improvements over the voxel-wise methods. However, the cross-sectional approach as implemented can not accommodate time-varying covariate effects, which results in poor performance compared to the proposed {\color{black}l-BTRR} method. Further, the improvements under the joint credible regions for feature selection {\color{black} under the tensor-based methods} is evident from higher F1 scores compared to the point-wise credible regions without multiplicity adjustments. This further points to the importance of using joint credible regions for feature selection, which is able to respect the shape of the posterior distribution. 

{\color{black} In terms of probabilities of coverage, we observe that the proposed l-BTRR method had the second highest coverage behind the longitudinal voxel-wise OLS method (vl-OLS) in all schemes after applying multiplicity corrections to obtain confidence intervals of prediction for the outcome. However, the width of the intervals used to obtain coverage probabilities was substantially lower for the l-BTRR method than the vl-OLS method. Coupled with the fact that the coverage probabilities for the l-BTRR method were always above 90\%, this finding illustrates the improved balance of accuracy and precision of the l-BTRR method over the voxel-wise competing approaches.}

While the MCMC sampling scheme tends to be more computationally expensive in practice compared to voxel-wise methods, our implementation is fairly efficient and runs within an hour for the example replicate we considered. Across all Schemes and replicates, average run-times were $445$s and $393$s per thousand iterations for l-BTRR and cs-BTRR, respectively, whereas faster run-times of $142$s, $4$s, $187$s, and $129$s were observed for vl-OLS, vcs-OLS, vl-Lasso, and vcs-Lasso, respectively.

\section{Aphasia Analysis }
\subsection{Data Description}
Fourteen subjects with post-stroke chronic aphasia (at least 6 months following stroke) were recruited for this study \citep{altmann2014delayed,krishnamurthy2021method, benjamin2014behavioral}.
After recording demographic information about subjects, including age ($67\pm 11$ years), gender (6 females, 8 males), and cerebrovascular accident type (11 ischemic and 3 hemorrhagic), each subject participated in both an MRI scanning session and task-fMRI language assessment session. For the MRI session, a 1 mm$^3$ isotropic high-resolution T1-weighted anatomical image for registration to Montreal Neurological Institute (MNI) template space was acquired using turbo field echo acquisition (echo time = $3.7$ ms, repetition time = $8.1$ ms, field of view = $240 \times 240$ mm$^2$, flip angle = 8$^{\circ}$, matrix size = $240 \times 240$). MRI scans were used to identify lesion volume and location for each subject. 
In tandem with MRI sessions, subjects underwent language assessment sessions. The language assessment included administration of the Western aphasia Battery - Revised \citep{kertesz2007western}, which generated an index of aphasia severity known as WAB-AQ.  Language treatment sessions included word retrieval probes to monitor progress, which consisted of both category-member generation and picture-naming trials. Subjects were randomly assigned to one of two treatment groups -- control ($n=7$) or intention ($n=7$). Both treatment groups underwent standard language therapy involving picture naming and category exemplar generation. In the intention group, an additional treatment involving complex left-hand motion during picture naming was administered, as described in \cite{benjamin2014behavioral}.

Task fMRI was used to survey brain activity during category-member generation over a total of three visits per subject (i.e. baseline, 2-week follow-up, and 3-month follow-up), except for one subject who dropped out of the study prior to the final follow-up visit. The task-fMRI scans involved a set of word retrieval tasks, where fMRI was used to survey brain activity during these tasks. In particular, for a total of 60 trials of 6.8 second duration each (6 runs of 10 trials), subjects heard and were shown text for a category (e.g. ``Tools'') and were instructed to speak a loud singular example of that category (e.g. ``Wrench''). Inter-trial intervals were of random duration between 13.6-17 seconds and consisted of subjects viewing a fixation cross while being instructed not to speak or move.



Task-based fMRI allows one to map treatment-induced brain reorganization and/or restoration when the person with aphasia (PWA) is engaged in a language task. One of the primary focuses of our study is to evaluate how brain neuroplasticity may vary with respect to the two treatments. The first step is to choose a suitable metric for capturing brain activity and associated neuroplasticity changes over longitudinal visits.
In this project, we use the voxel-wise area under the curve (AUC) to measure brain activity induced by the task experiment and compute the longitudinal AUC differences across visits to evaluate neuroplasticity, which is more robust to noise \citep{krishnamurthy2020correcting}. The AUC is an integration of percent change BOLD activity underneath the estimated hemodynamic response function for a given voxel. It is agnostic to treatment-specific and session-specific variability in peak amplitude changes, which is desirable given that estimates of peak amplitude can be heavily influenced by false-positive artifacts, fMRI properties (e.g. sampling rate), and modeling assumptions (e.g. not adequately accounting for temporal variation) 
\citep{miezin2000characterizing, lindquist2009modeling}. From a physiological and biophysical viewpoint, AUC is a good marker for task-induced brain energetics and thereby is suitable for evaluating treatment-induced neuroplasticity changes \citep{krishnamurthy2020correcting}. 


\vskip 10pt

\noindent \underline{Screening out missing voxels}: In practice, each subject is expected to have a subset of voxels in the brain image that are treated as `missing' or redundant, especially in stroke studies. These voxels are considered redundant due to the fact {\color{black} that (i) they lie outside of the brain mask;} (ii) they belong to the lesion areas with disrupted brain activity and hence are not expected to show neural plasticity changes that are of primary interest; or (iii) they record zero or close to zero brain activity in terms of AUC values across all samples (i.e., they show no evidence of hemodynamic response activity in any subject at any time) and hence are not discriminatory for our analysis. The set of redundant voxels in (ii) is expected to vary across individuals depending on the lesion characteristics, while the set of voxels in (iii) is common across all samples. We implement a screening step to exclude these redundant voxels, which is a practical step that leads to a dimension reduction for the outcome image used without loss of accuracy. {\color{black} 
This screening step also makes the tensor-based methods more comparable to the univariate voxel-wise analysis, which requires these voxels to be excluded from analysis, since they either have zero AUC values across all or most samples, preventing a reasonable effort to fit a model corresponding to these voxels.} 
After screening, the number of remaining voxels for analysis had a range of 26469-30200 across the 14 study individuals. Furthermore, all subjects had AUC scores for three clinical visits -- baseline ($t=0, \ \mathcal{T}_{0i}=0\text{ days}$), post-treatment ($t=1, \ \mathcal{T}_{1i}\approx49 \text{ days}$), and 3-month follow-up ($t=2, \ \mathcal{T}_{2i}\approx97 \text{ days}$)-- except for one subject who dropped out after Visit $t=1$.



\subsection{Model for Aphasia Study} 
Prior to model fitting, all continuous covariates were standardized across the 14 subjects using z-scores. The z-transformed AUC score serves as the tensor response in our analysis and is denoted as $\mathcal{Y} \in \mathbb{R}^{45\times 58 \times 49}$. Additional covariates include age, gender, lesion volume (Lesvol), and aphasia severity (WAB-AQ), {\color{black}along with treatment (1=intention treatment, 0=control treatment), where `Trt' denotes the binary treatment variable.} Our goals for this analysis include inferring voxels showing significant neuroplasticity changes between visits, stratified between different levels of covariates. 
An increase or decrease in AUC values between consecutive visits indicates neuroplasticity changes. We formulate the following model to address the above goals under a unified framework:

\begin{eqnarray}
&& y_{ti}(v) = \mathcal{M}(v) + B_i(v) + \Gamma(v) \mathcal{T}_{ti}  + \Theta_i(v) \mathcal{T}_{ti} + \mathcal{D}_1(v) \text{Age}_i + \mathcal{D}_2(v) \text{Gender}_i + \mathcal{D}_3(v) \text{Lesvol}_i  \nonumber \\ 
&& \text{ } \text{ } \text{ } \text{ } \text{ } \text{ } \text{ } \text{ } \text{ } \text{ } \text{ } \text{ } \text{ } \text{ } \text{ } \text{ } \text{ } \text{ } \text{ } + \mathcal{D}_4(v) \text{WAB-AQ}_i + {\color{black} \mathcal{C}_1(v)1(\text{Trt}_i=0) 1(t=2)} + \mathcal{B}_{t}(v) 1(\text{Trt}_i=1) + \epsilon_{ti}, \mbox{ }  v\in\mathcal{V}_i, \mbox{ } \label{eq:modelaphasia}
\end{eqnarray}

\noindent where $\mathcal{B}_0(v)=0$, $y_{ti}(v)$ is the AUC score for subject $i$, visit $t$, and voxel $v$, $\mathcal{M},  B_i, \Theta_i$, and $\Gamma$ are defined as in equation (\ref{eq:longBTRR}), $\mathcal{D}_1,\mathcal{D}_2,$ are demographic-related covariate effects, $\mathcal{D}_3,\mathcal{D}_4,$ are effects of covariates related to the severity of the disease, and {\color{black} $\mathcal{C}_1$}, $\mathcal{B}_1,\mathcal{B}_2,$ represent group-level treatment effects that impact neuroplasticity scores. In particular, 
the coefficients {\color{black} $(\Gamma + \Theta_i)\mathcal{T}_{1i}$ and $(\Gamma + \Theta_i)(\mathcal{T}_{2i}-\mathcal{T}_{1i})+ \mathcal{C}_1 $} capture the group-level neuroplasticity between baseline and visit 1, and visits 1 and 2 respectively, corresponding to the $i$th subject in the control group. Similarly, {\color{black} $(\Gamma + \Theta_i)\mathcal{T}_{1i}+ \mathcal{B}_1(v)$ and $(\Gamma + \Theta_i)(\mathcal{T}_{2i}-\mathcal{T}_{1i}) + (\mathcal{B}_2(v)-\mathcal{B}_1(v))$} capture the neuroplasticity changes between baseline and visit 1, and between visits 1 and 2, respectively, corresponding to the $i$th subject in the intention treatment group. 
Finally, the noise term $\epsilon_{ti}(v)$ is assumed to be Normally distributed with mean 0. We note that the proposed model is flexible in allowing for differential neuroplasticity changes related to treatment across different pairs of time visits, while also allowing for subject level heterogeneity. Further, even with low sample sizes, high voxel counts, and relatively large numbers of covariates, the proposed tensor approach can produce robust parameter estimates by borrowing information across voxels via low-rank decomposition.


\subsection{Results}

The results below report group-level as well as individual neuroplasticity changes under the proposed tensor model,  which are inferred using the joint credible intervals approach. 
We note that we also performed a voxel-wise regression for the aphasia dataset. {\color{black} We observed that using this approach, there was a large proportion of voxels for which the full model including all covariates and subject-specific terms did not converge. For these voxels, we therefore had to refit the voxel-wise model by excluding the subject-specific time-slope. However, this} approach did not yield any significant neuroplasticity changes due to treatment after {\color{black} CEI-based} multiplicity adjustments, and therefore no voxel-wise results are presented under this analysis. Such results are biologically implausible and point to the challenges under the voxel-wise analysis, as highlighted in the Introduction and Section 2.


\subsubsection{Group-level Neuroplasticity Maps}

Our focus is to evaluate the differences in neuroplasticity maps between different levels of covariates included in the model, based on the approach presented in Section 2.4. This {\color{black} stratification} strategy is designed to {\color{black} investigate the impact of clinical factors on neuroplasticity.}  We present the neuroplasticity maps between baseline and the post-treatment visit, as well as between the post-treatment and three-month follow-up visits, corresponding to (i) the two treatment groups; (ii) two age groups ($<65$ and $\ge 65$ years); (iii) moderate and mild aphasia severity (WAB-AQ score between 50-75 and 75-93.8 respectively); and (iv) varying levels of lesion volumes (lower, middle, and upper tertile). 


We start with overall neuroplasticity maps that represented the group-level changes without any stratification based on covariates  (top panel of Figure \ref{fig:Figure4}). For the overall neuroplasticity involving all fourteen subjects, decreased activity was observed within the right middle temporal gyrus (R-MTG), and increased activity was observed within the right precuneus between baseline to post-treatment. In contrast, only activity increases were observed within the right angular gyrus (R-AG) and left middle frontal gyrus (L-MFG) between post-treatment and 3-month follow-up ({\color{black}Figure \ref{fig:Figure4}}) 

\begin{figure}[H]
\centering
\begin{subfigure}{0.9\textwidth}
  \centering
  \includegraphics[width=\textwidth]{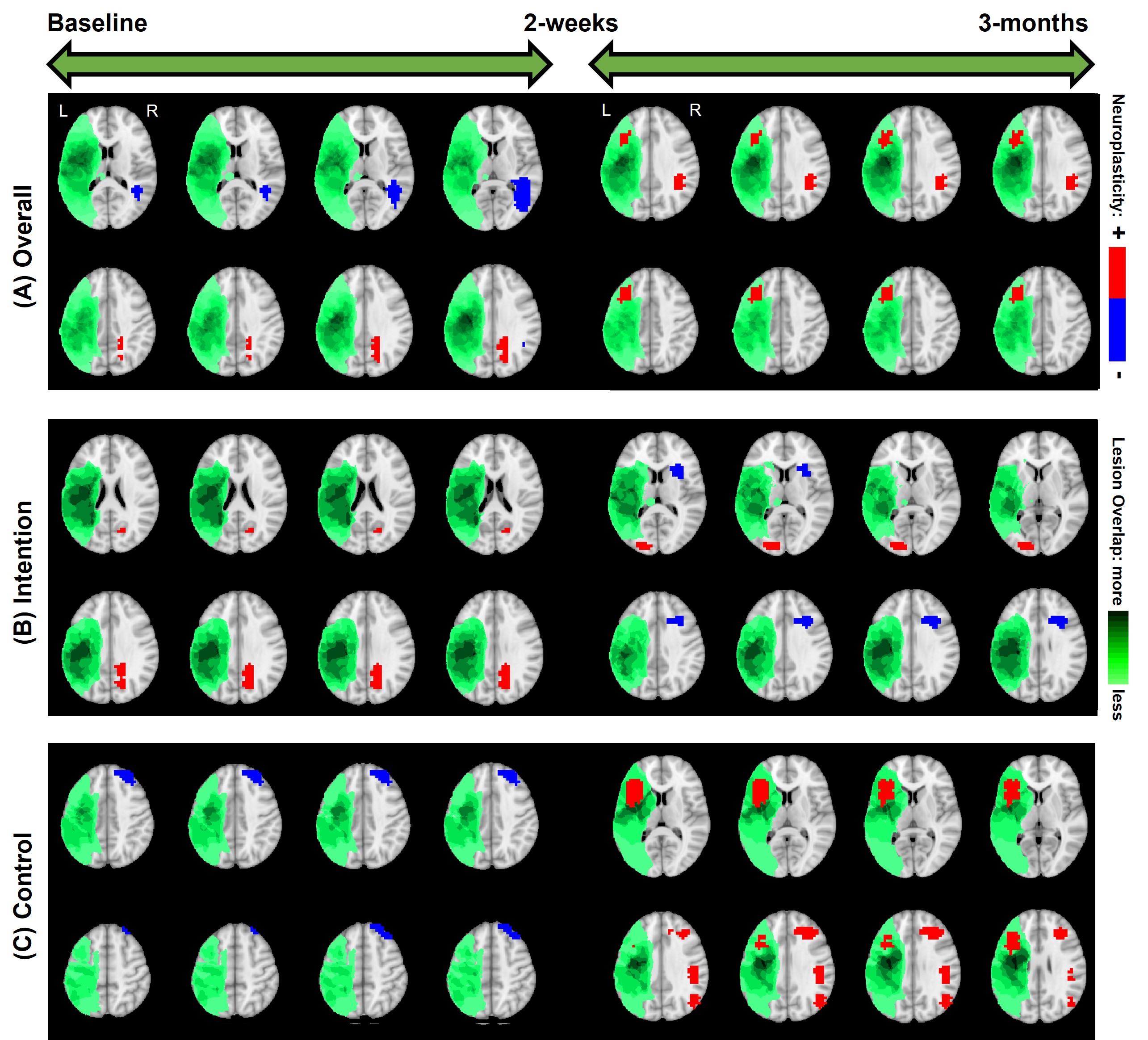}
\end{subfigure}
\caption{\label{fig:Figure4} \textbf{Overall and treatment-specific neuroplasticity.} Significance estimates for neuroplasticity across all subjects (top), intention group (middle), and control group (bottom) from baseline to 2-week visit and 2-week to 3-month visit. Each 3D estimate is portrayed as a set of eight 2D brain slices with the largest amount of significant voxels, where regions of significant increased (+) and decreased (-) brain activity are overlayed with a heatmap of lesion locations.}
\end{figure}

\vskip 10pt

{\noindent \underline{Maps stratified by treatment:}}  Within the intention treatment group, the right precuneus displayed increased activity between baseline to 2 weeks post-treatment while the {\color{black}right inferior frontal gyrus (R-IFG)}, specifically pars opercularis, displayed decreased activity while the {\color{black}left middle occipital gyrus (L-MOG)} displayed increased activity between 2 weeks and 3 months post-treatment ({\color{black}middle and bottom panels of Figure \ref{fig:Figure4}}). Within the  control group, the right superior frontal gyrus (R-SFG) displayed decreased activity between baseline to 2 weeks post-treatment while the L-IFG (pars opercularis), {\color{black}right middle frontal gyrus (R-MFG)}, right supramarginal gyrus (R-SMG), and R-MOG displayed increased activity between the 2 weeks and 3 months post-treatment ({\color{black}Figure \ref{fig:Figure4}}). {\color{black}Both the overall neuroplasticity maps (three months post baseline) and the control treatment maps show long-term activity increases in pericavitational and perilesional brain areas.}

\vskip 10pt

{\noindent \underline{Maps stratified by age:}} 
In terms of age ({\color{black} Figure \ref{fig:Figure5}}), for participants $< 65$ years old, the right superior temporal gyrus (R-STG) decreased activity from baseline to post-treatment while bilateral middle frontal gyrus (L \& R-MFG) and R-STG showed increased activity from  post-treatment to 3-month follow-up. For participants $\ge$ 65 years old, the R-MTG showed decreased activity from baseline to post-treatment while the left inferior frontal gyrus (L-IFG) showed increased activity from post-treatment to follow-up. In contrast, the contralateral R-IFG showed decreased activity from post-treatment to 3-month follow-up. Overall, according to \cite{ellis2016age}, participants younger than 65 have more rehabilitation potential to benefit from treatment-specific plastic changes, and those older than 65 may need more tailored and additional treatments to recover long-term increased brain activity.

\vskip 10pt
{\noindent \underline{Maps stratified by aphasia severity:}} 
For the moderate aphasia severity group, the {\color{black} right inferior frontal gyrus (R-IFG)} negatively influenced short-term neuroplasticity, while the contralateral L-IFG and R-MTG positively influenced long-term neuroplasticity. On the other hand, participants in the mild aphasia severity group displayed significant clusters in the {\color{black} right superior temporal gyrus (R-STG)} that negatively influenced short-term neuroplasticity and some subcortical areas such as the right caudate and putamen that positively influenced short-term neuroplasticity. In terms of long-term neuroplasticity, the right middle occipital gyrus (R-MOG) was found to have a positive influence while the R-MFG was found to have a negative influence. {\color{black} The moderate severity group exhibited long-term increased neuroplasticity changes in pericavitational and perilesional brain areas. } These results are visually illustrated in {\color{black}Figure \ref{fig:Figure5}}.

\vskip 10pt
{\noindent \underline{Maps stratified by lesion volume:}} 
In terms of the lesion volume, participants within the lower 1/3rd quantile displayed significant clusters in the R-MFG that decreased activity from baseline to post-treatment while both the L-IFG and R-MTG increased activity from post-treatment to 3-month follow-up. Participants within the middle 1/3rd quantile displayed significant clusters in the R-MOG that decreased activity from baseline to post-treatment while the right lingual gyrus (R-LG) increased activity from post-treatment to 3-month follow-up. Participants within the upper 1/3rd quantile displayed significant clusters in the right precentral gyrus that decreased activity from pre-treatment to follow-up while the left insula and R-LG increased activity from post-treatment to 3-month follow-up. The group with higher lesion volume showed long-term activity increases in brain regions lying close to the lesion areas. These results are visually illustrated in  {\color{black}Figure \ref{fig:Figure5}.}




\vskip 10pt
{\noindent \underline{Summary Comments}}: It is interesting to observe that the model provided consistently increased brain activity estimates for long-term changes when all participants were pooled together irrespective of specific (i.e. standard or intention) therapy. Further, when the participants were separated based on the type of treatment, our novel modeling approach was able to identify unique {\color{black} potential} biomarkers for treatment-specific neuroplasticity changes. While the control therapy showed a long-term activity increase, the intention treatment provoked both short- and long-term activity increases. On the other hand, activity decreases were in short-term measures for the control treatment and only in long-term measures for the intention treatment. Considering that intention treatment involved additional non-gestural circular hand movements on top of the standard therapy, we hypothesize that such cognitively (i.e. intention) driven non-symbolic hand movements facilitate cognitive control and gating of information flow \citep{gratton2018brain}.

\begin{figure}[H]
\centering
\begin{subfigure}{0.90\textwidth}
  \centering
  \includegraphics[width=\textwidth]{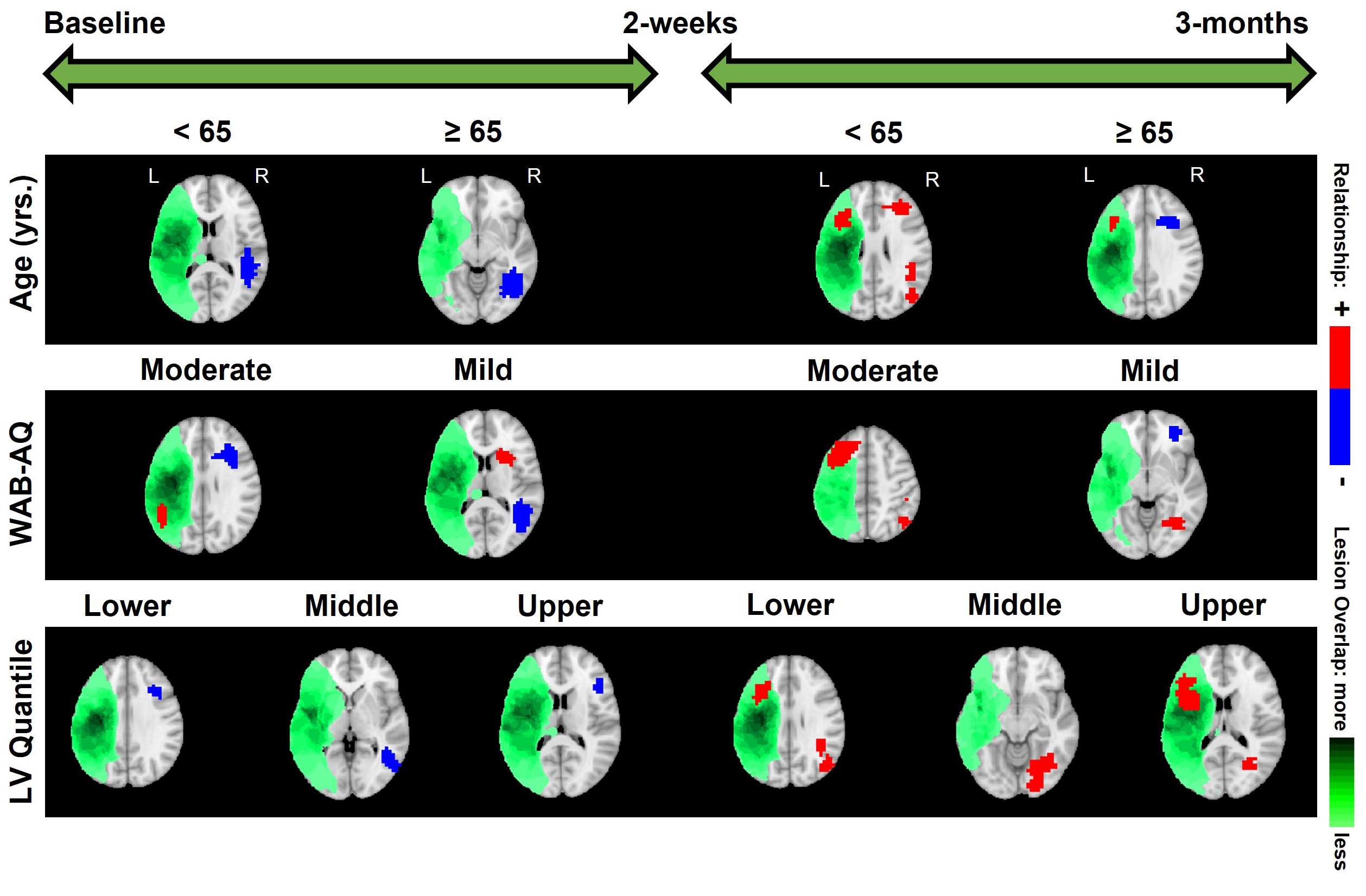}
\end{subfigure}
\caption{\label{fig:Figure5} \textbf{Neuroplasticity by age, severity, and lesion volume}. Significance estimates for group-specific neuroplasticity corresponding to age, WAB-AQ, and lesion volume (LV), shown for each consecutive follow-up. Regions of significant increased (+) and decreased (-) brain activity are overlayed with a heatmap of lesion locations for the corresponding covariate group.}
\end{figure}

\subsubsection{Individual neuroplasticity maps:} In addition to the above group-level findings, our novel model also allows to extract neuroplasticity maps for each individual over visits. Common markers of short-term neuroplasticity within the intention group participants included the R-IFG (pars triangularis) for increased activity changes and right inferior temporal gyrus (R-ITG) for decreased activity changes whereas long-term neuroplasticity estimates were more common within the R-MFG for decreased activity and the R/L-MOG for increased activity ({\color{black}Figure \ref{fig:Figure6}}). Other, less common, significant clusters were found within right hemisphere subcortical regions corresponding to increased short-term activity as well as decreased long-term activity changes. Common markers of short-term neuroplasticity within the control group participants included the L-MOG and R/L-MFG for decreased activity changes whereas long-term neuroplasticity estimates were more common within the R-MFG and L-IFG (pars triangularis) for increased activity changes (Figure \ref{fig:Figure6}). 
Considering the heterogeneity of the lesion profile and associated post-stroke recovery, generating personalized diagnostic and prognostic biomarkers are very pragmatic and attractive in stroke rehabilitation. Our results not only show that our novel approach has the potential to generate such individualized maps, but also the individualized results show consistent trends in neuroplasticity changes that are treatment and time point specific. These personalized treatment-specific spatial maps of neuroplasticity also allow for potential triaging of participants into individualized treatment plans that are tailored to their baseline clinical profiles. 

\begin{figure}[H]
\centering
\begin{subfigure}{0.49\textwidth}
  \centering
  \includegraphics[width=\textwidth]{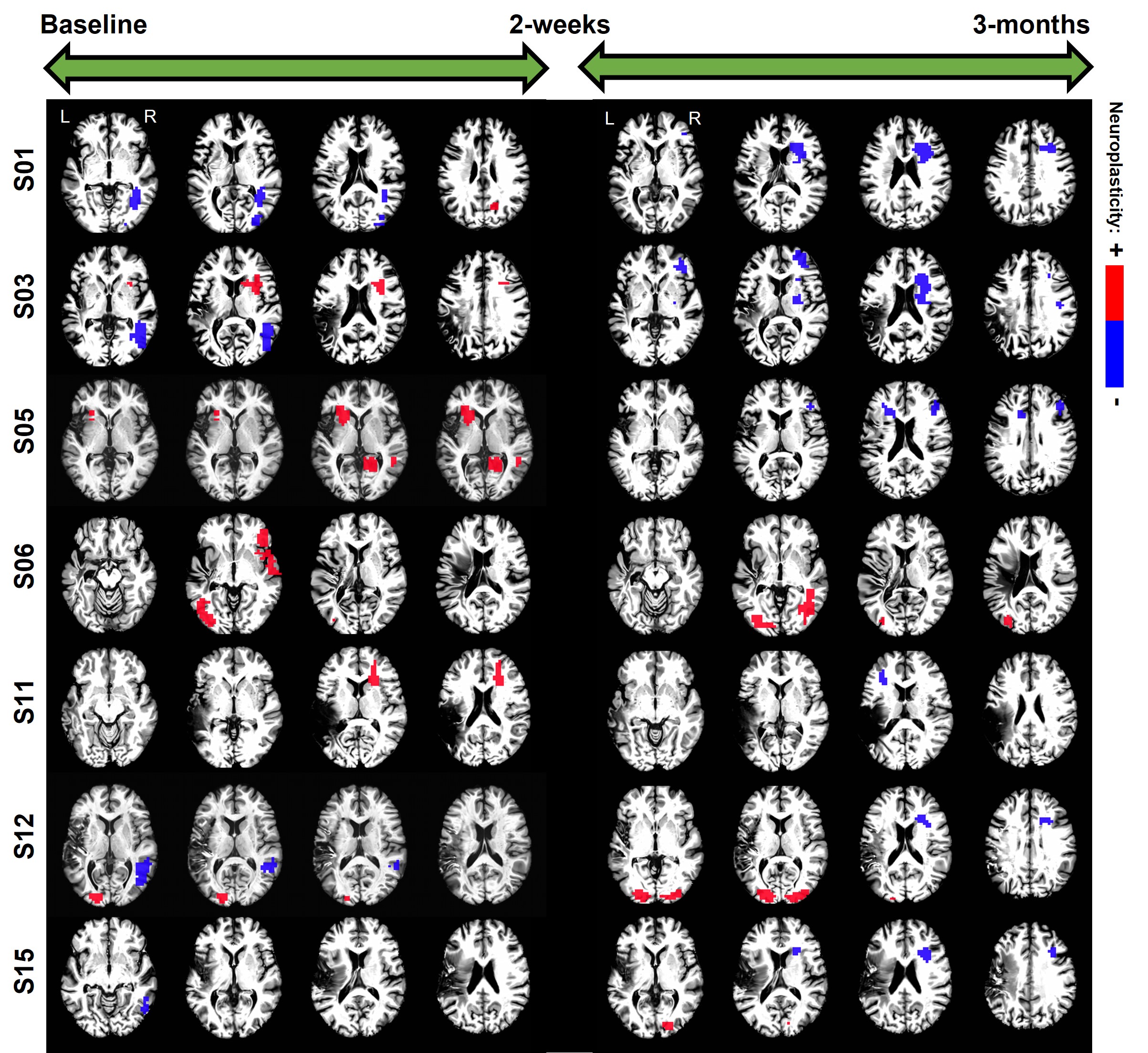}
  \caption{}
\end{subfigure}
\hfill
\begin{subfigure}{0.49\textwidth}
  \centering
  \includegraphics[width=\textwidth]{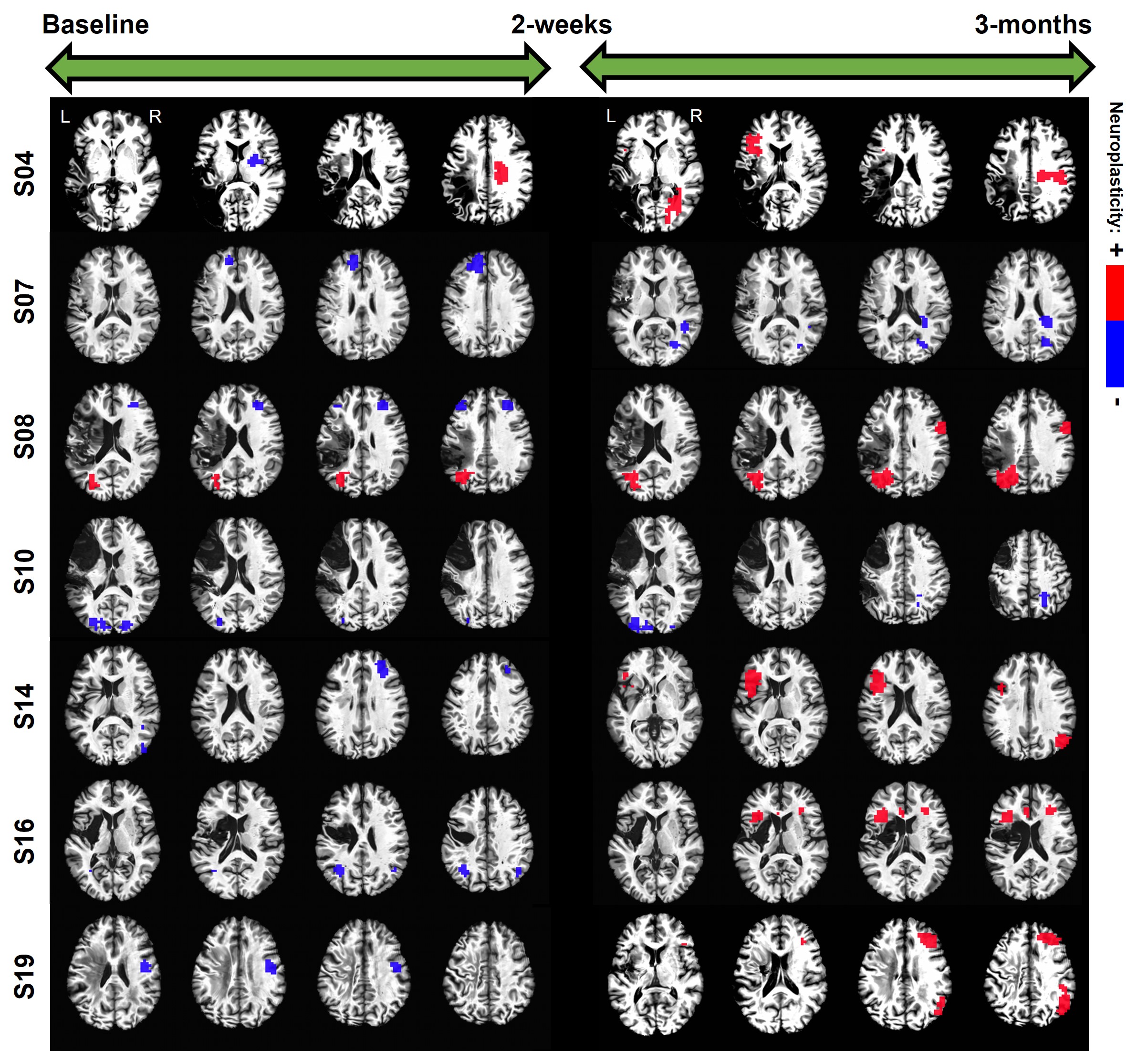}
  \caption{}
\end{subfigure}
\caption{\label{fig:Figure6} \textbf{Subject-specific neuroplasticity maps}. Significance estimates for individual level neuroplasticity among (a). subjects in intention therapy group, and (b). subjects in the {\color{black} control} therapy group. Regions of significant increased (+) and decreased (-) brain activity are shown for each subject across baseline to 2-weeks and 2-weeks to 3-months.}
\end{figure}

\subsubsection{Validation Using Longitudinal Prediction}

To assess the suitability of the tensor model for the aphasia data, we evaluated the out-of-sample prediction performance of the proposed model and other competing methods  (cs-BTRR, vl-OLS, vcs-OLS, vl-Lasso). The training set comprised all voxels from the first two visits for all 14 subjects, whereas the test set comprised varying levels of holdout voxels from the third visit. In particular, we considered randomly chosen $20\%$, $40\%$, 50\% and $60\%$ of non-missing voxels pertaining to Visit 2 across all subjects, to construct the test set. For each holdout level and method, the out-of-sample root mean squared error (RMSE) was computed to examine predictive performance. For the l-BTRR method, 5 different choices of ranks $R$ were used (i.e. $R=1,\ldots,5$) to ascertain the effect of rank on predictive performance. As seen in Figure \ref{fig:Figure7}a, l-BTRR (using rank 3) had superior out-of-sample predictive accuracy compared to all competing methods for low to moderate holdout (i.e. roughly less than 60\%), but the performance becomes comparable to the cross-sectional BTRR for higher levels of holdout voxels. These findings validate the prowess of the longitudinal BTRR approach over both cross-sectional BTRR and voxel-wise methods when the number of missing voxels is a small to moderate.  Moreover, the voxel-wise methods have consistently inferior predictive accuracy compared to tensor-based approaches that is not surprising given our simulation findings.   

Figure \ref{fig:Figure7}b shows the out-of-sample RMSE  for l-BTRR with a choice of rank between 1-5, demonstrating how higher choices of rank have better predictive performance for low holdouts but are more sensitive to the size of the training set and perform worse on high holdouts. This is expected, given that tensor models with higher ranks contain more parameters and require larger sample sizes for optimal training. To further assess the optimal choice of rank, the DIC score was computed for each level of holdout (20\%-60\%) and rank (1-5). We found that the rank 3 model fits yielded a lower DIC than all other examined ranks across all levels of holdout. Given these results, rank 3 was selected for the full aphasia analysis, as it provided a balance of model parsimony and predictive performance.

\begin{figure}[H]
\centering
\begin{subfigure}{.45\textwidth}
  \centering
  \includegraphics[width=\textwidth]{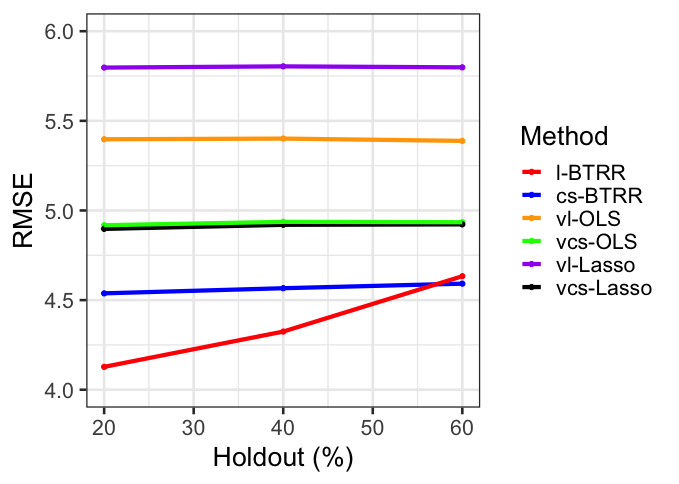}
  \caption{}
\end{subfigure}
\begin{subfigure}{.45\textwidth}
  \centering
  \includegraphics[width=\textwidth]{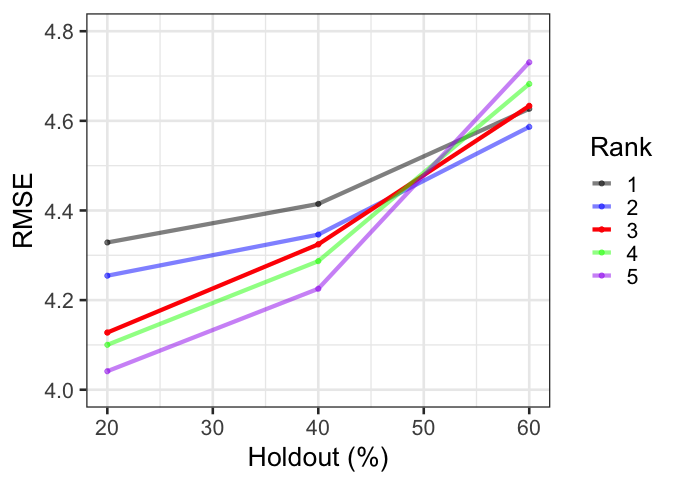}
  \caption{}
\end{subfigure}
\caption{\textbf{Model evaluation on aphasia data}. Out-of-sample RMSEs across 3 levels of holdout on Visit 2 for (a). rank-3 l-BTRR, rank-3 cs-BTRR, and 4 competing voxel-wise methods, and (b). l-BTRR using ranks 1-5}
\label{fig:Figure7}
\end{figure}

\subsubsection{Model Diagnostics}

{\color{black} In order to further evaluate the performance and biological validity of the proposed l-BTRR when fitting to the aphasia dataset, we examined various other criteria. In particular, we looked at convergence diagnostics of the MCMC samples for the estimated model coefficients and fitted AUC outcome in Model \ref{eq:modelaphasia} at each voxel. Figure \ref{fig:Figure8} provides example traceplots of the fitted AUC outcome (using the l-BTRR method) across 2500 post-burn-in MCMC samples for two randomly-selected voxels, one with significant (non-zero) fitted AUC and one with non-significant (zero) fitted AUC. These traceplots provide visual evidence for convergence in the MCMC chain under the l-BTRR approach. To test for convergence more rigorously, we also ran the Dickey-Fuller test \citep{dickey1979distribution} on each post-burn-in MCMC chain of the fitted AUC outcome, averaged across all subjects and time points. Of all the traceplots for the 30200 non-missing voxels, $26425$ ($87.5$\%) voxels had corresponding Dickey-Fuller test p-values of less than $0.05$, providing evidence that these voxels were stationary. In comparison, we performed the same set of tests on the cs-BTRR model fit to the aphasia dataset and found that $27572$ voxels ($91.3\%$) had evidence of stationarity under the cross-sectional tensor framework. The convergence rate could potentially be increased further by increasing the number of MCMC samples. Moreover, we note that simplifying the proposed l-BTRR approach by omitting model terms and fitting it as a cross-sectional model increases the efficiency of the MCMC sampler, leading to higher rates of convergence across voxels. However, the increased efficiency of the cs-BTRR MCMC algorithm comes at the cost of losing capability for individual-level inference and having worse estimation and out-of-sample prediction. Given the fact that the number of voxels with stationary traceplots did not differ drastically across methods, we conclude that the advantages of the l-BTRR method outweigh the costs in this application. The mixing of the MCMC chain also depends on additional tuning parameters, such as $\sigma_{\alpha}^2$, which is involved in the Metropolis-Hastings step.}

\begin{figure}[H]
    \centering
    \includegraphics[width=\textwidth]{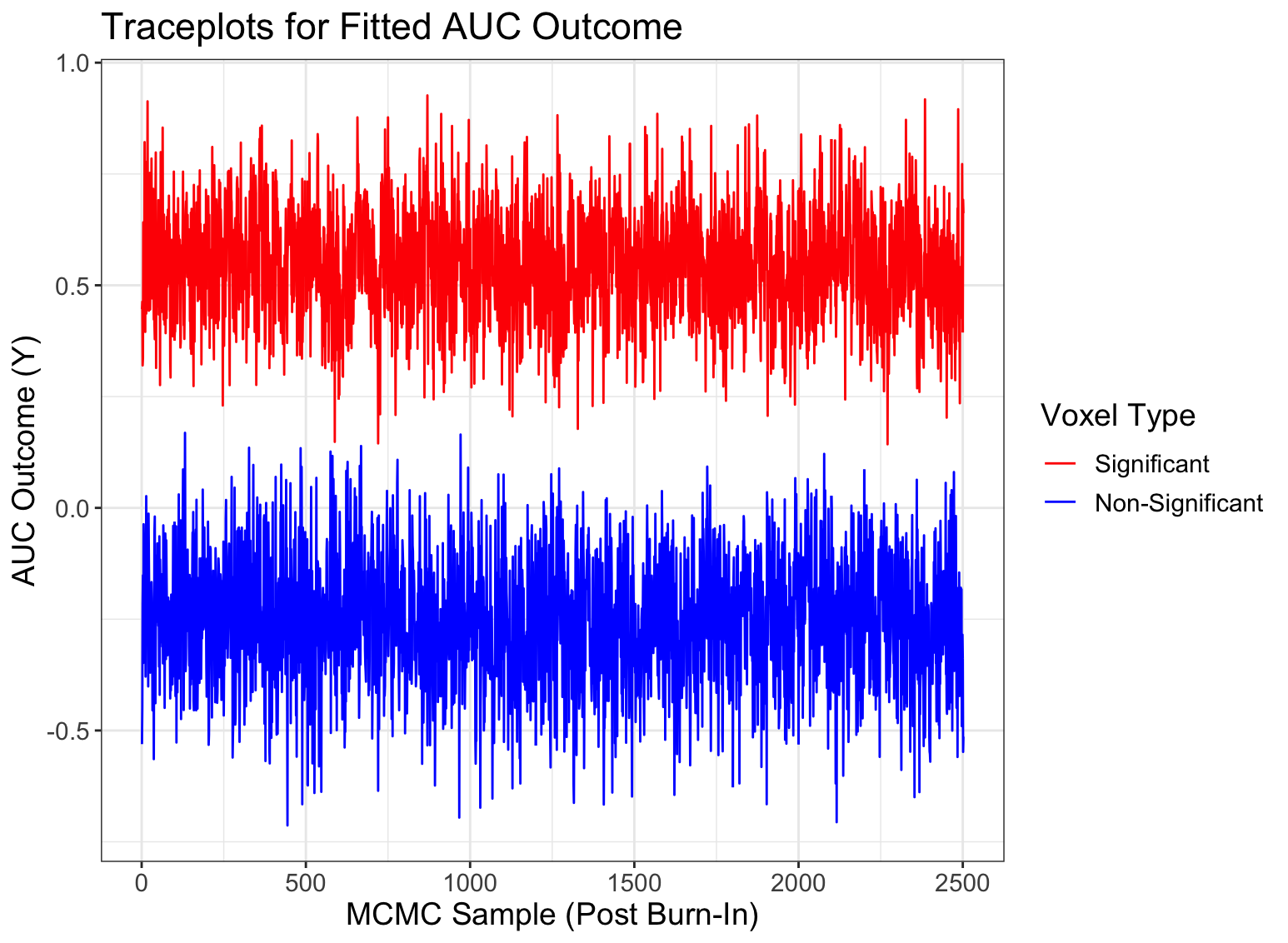}
    \caption{Example traceplots for fitted AUC outcome averaged across subjects in Aphasia analysis. The blue curve corresponds to a randomly selected voxel that had a fitted AUC not significantly different than 0, and the red curve corresponds to a voxel with significant (non-zero) fitted AUC outcome. }
    \label{fig:Figure8}
\end{figure}

\section{Discussion}

Our analysis of data from an aphasia longitudinal neuroimaging study revealed distinct spatial patterns of neuroplasticity that may vary by treatment and/or with respect to clinical covariates. We conclude that neuroplasticity changes may consist of activity increases or decreases depending on treatment. 
These general findings, that brain functionality is heterogeneous across space, time, and subject, are consistent with the literature.  It is also worthy of comment that long-term changes in brain activity occurred between post-treatment and 3-month follow-up fMRI. Such long-term changes occurred for both the intention and control treatments but were more pronounced in the control group. Long-term brain activity changes continuing 3 months beyond treatment might reflect the improvement in discourse output. In the short-term (post-treatment), the intention group resulted in brain activity changes whereas the control group led to decreases in brain activity. Thus, the pattern of short-term and long-term brain activity changes are fairly distinct between the treatment groups, which may potentially contribute to differences in discourse gains. Additional work is needed for a more systematic interpretation to determine the clinical significance of those brain regions that are most highly affected.




One must be careful when interpreting the meaning of the neuroimaging results from the aphasia study. Since the goal of rehabilitation is to change functional behaviors, the ultimate determination of whether changes in brain activity are adaptive, maladaptive, or neutral depends on whether they are associated with improvement in target rehabilitation language behaviors. This study focused on improving analysis techniques to determine more accurately the brain regions in which activity increased or decreased. Therefore, the implications of these changes in the light of modifications in language behaviors were beyond the scope of this paper and will be considered in future work. Indeed, there is previous evidence of association between language behavior and changes in brain activity. In particular, in an analysis based on a voxel-wise approach, a rightward shift in posterior perisylvian activity from baseline to post-treatment was found to be associated with improved word retrieval only for the intention but not for the control group \citep{benjamin2014behavioral}. Hence, the next step in our research is to incorporate language outcome data into the analysis models for neuroplasticity under the elegant framework of Bayesian tensor modeling. Such analysis will enable us to confirm/disconfirm  previous findings and potentially shed light on new findings that were not discovered under a voxel-wise approach, likely due to the limitations outlined in this article.

While it is worthwhile to identify that lesion volume influences long-term plasticity, our results indicate that age is a critical factor in producing increases in brain activity. This is consistent with the observations in \cite{ellis2016age} that participants younger than 65 have more rehabilitation potential to benefit from treatment-specific plastic changes, and those older than 65 may need more tailored and additional treatments to gain increased brain function in the long-term.  Modeling clinical factors such as age, aphasia severity, and lesion volume brought out activity decreases from baseline to post-treatment.  This outcome is not negative, per se. As pointed out above, its clinical significance depends on whether it is associated with desired changes in language behavior. Further, since these analyses were collapsed across the competing treatments, it is unclear how much influence each treatment had on these findings. Additionally, one emerging principle pertaining to short-term synaptic plasticity is that changes in the balance of parallel excitation and feedforward inhibition can be used for gating information flow in an activity-dependent manner \citep{anwar2017functional, bao2010target}. Since excitation in a structure would tend to increase neural activity while the results of feedforward inhibition in the same structure would tend to decrease activity, this phenomenon represents a challenge for visualization and quantification using advanced neuroimaging tools. Yet, understanding these phenomena could be useful in treatment planning tailored to a participant’s baseline clinical profile. This is particularly relevant given the ability to up- or down-regulate focal brain activity with non-invasive brain stimulation techniques, which is emerging as a promising technique in aphasia treatment (see review in \cite{crosson2019neuroplasticity}). The subject-specific neuroplasticity maps inferred under the proposed method can potentially serve as an important tool for determining such personalized treatment interventions. {\color{black}Finally, we note that the proposed tensor-based approach succeeds in providing accurate results for the aphasia study with a small sample size, which is encouraging and highlights the robustness of the method. We note that such small sample sizes are routinely encountered in stroke as well as brain tumor literature, which are rare disorders. The proposed approach can be generalized to other studies with image outcomes and larger sample sizes in a straightforward manner.}

{\color{black}The proposed tensor-based approach involves a careful selection of the tensor rank using the DIC score, which is routinely used in literature. However, this strategy requires us to fit separate models for each choice of rank which can be potentially time-consuming. A possible alternative is to perform trans-dimensional MCMC where the tensor rank is learned in an adaptive manner, which would also provide uncertainty estimates about the tensor rank. This method could be potentially explored in future work. A limitation of most tensor-based approaches is that they are expected to deteriorate when the level of discontinuities becomes increasingly
high or where there are sharp localized discontinuities (e.g. sparse cortical thickness images). Such applications may need higher choices of tensor rank or more specialized choices of basis functions such as wavelets.}

\section{Conclusion}

In this article, we developed a novel {\color{black}and scalable tensor response regression approach that models} voxel-level brain imaging features {\color{black}based} on covariates of interest for longitudinal neuroimaging studies, which is designed to provide considerable advantages over routinely used voxel-wise approaches. The proposed approach is able to preserve the spatial configurations of the voxels, it accommodates heterogeneity between samples while also allowing for group-level inferences, and it is able to pool information across voxels, yielding model parsimony and improved feature selection accuracy.  The importance of the tensor-based approach for the analysis of aphasia data becomes evident from superior out-of-sample predictive performance over voxel-wise methods, and given the fact that the voxel-wise approach is unable to infer any significant neuroplasticity changes after multiplicity adjustments. {\color{black}The proposed approach should be applicable to a wide array of longitudinal imaging studies and can be used in lieu of voxel-wise analysis to produce more accurate results.} 

\printbibliography[heading=bibintoc,title={References}]

\vskip 10pt

\begin{table}[H]
    \centering
    \begin{tabular}{|c|c|c|c|c|c|c|c|}
        \toprule
        \multicolumn{8}{|c|}{Holdout 25\%} \\
        \midrule
        Method & p-RMSE & p-Corr & c-RMSE & Sens & Spec & F1 & Coverage \\
        \midrule
         l-BTRR & \textbf{1.706} & \textbf{0.681} & \textbf{0.082} & 0.951 & 0.995 & \textbf{0.965}  & 0.969 (1.722) \\
         cs-BTRR & 1.861 & 0.620 & 0.119 & \textbf{1.000} & 0.945 & 0.931  & 0.616 (\textbf{1.050}) \\
         vl-OLS & 2.167 & 0.519 & 0.343 & 0.404 & \textbf{0.999} & 0.572  & \textbf{0.994} (4.810) \\
         vcs-OLS & 2.075 & 0.504 & 0.343 & 0.404 & \textbf{0.999}  & 0.572  & 0.845 (2.419) \\
        vl-Lasso & 2.179 & 0.528 & 0.345 & 0.979 & 0.167 & 0.415 & - (-) \\
        vcs-Lasso & 2.063 & 0.501 & 0.342 & 0.834 & 0.479 & 0.524 & - (-) \\
        \midrule
        \multicolumn{8}{|c|}{Holdout 50\%} \\
        \midrule
         l-BTRR & \textbf{1.713} & \textbf{0.678} & \textbf{0.086} & 0.949 & 0.996 & \textbf{0.965}  & 0.959 (1.704) \\
         cs-BTRR & 1.888 & 0.609 & 0.119 & \textbf{1.000} & 0.924 & 0.911 & 0.578 (\textbf{0.964}) \\
         vl-OLS & 2.167 & 0.519 & 0.364 & 0.361 & \textbf{0.999} & 0.527  & \textbf{0.994} (4.808) \\
         vcs-OLS & 2.077 & 0.504 & 0.362 & 0.361 & \textbf{0.999} & 0.527  & 0.845 (2.417) \\
         vl-Lasso & 2.183 & 0.526 & 0.364 & 0.959 & 0.164 & 0.403 & - (-) \\
        vcs-Lasso & 2.065 & 0.498 & 0.361 & 0.707  & 0.480  & 0.491   & - (-) \\
        \bottomrule
    \end{tabular}
    \caption{{\bf Scheme 1 Results}. Out-of-sample predictive performance (p-RMSE and p-Corr), estimated vs. true coefficient correlation across voxels (c-RMSE), feature selection (Sens, Spec, F1), and coverage probability (interval width in parentheses) for 6 competing methods. For feature selection, metrics are shown with multiplicity correction, where joint credible intervals are used for tensor-based approaches and cluster-extent inference is used for voxel-wise OLS.}
    \label{tab:Table3}
\end{table}

\begin{table}[H]
    \centering
    \begin{tabular}{|c|c|c|c|c|c|c|c|}
        \toprule
        \multicolumn{8}{|c|}{Sphere: Holdout 25\%} \\
        \midrule
        Method & p-RMSE & p-Corr & c-RMSE & Sens & Spec & F1 & Coverage \\
        \midrule
         l-BTRR & \textbf{1.735} & \textbf{0.673} & \textbf{0.154} & 0.966 & 0.956 & \textbf{0.919}  & 0.936 (2.306) \\
         cs-BTRR & 1.822 & 0.645 & 0.161 & \textbf{0.991} & 0.890  & 0.865   & 0.682 (\textbf{0.995})  \\
         vl-OLS & 2.140 & 0.539 & 0.339 & 0.315 & \textbf{0.999} & 0.471  & \textbf{0.995} (4.729) \\
         vcs-OLS & 2.006 & 0.545 & 0.338  & 0.300  & \textbf{0.999} & 0.456  & 0.874 (2.357)  \\
         vl-Lasso & 2.152 & 0.543 & 0.343  & 0.981  & 0.172  & 0.491   & - (-) \\
        vcs-Lasso & 1.997 & 0.538 & 0.336  & 0.861  & 0.591  & 0.596  & - (-) \\
        \midrule
        \multicolumn{8}{|c|}{Sphere: Holdout 50\%} \\
        \midrule
         l-BTRR & \textbf{1.742} & \textbf{0.671} & \textbf{0.156} & 0.959 & 0.951 & \textbf{0.907}  & 0.911 (1.899) \\
         cs-BTRR & 1.841 & 0.638 & 0.162 & \textbf{0.991} & 0.882  & 0.859   & 0.682 (\textbf{0.993})  \\
         vl-OLS & 2.140 & 0.537 & 0.360 & 0.295 & \textbf{0.999} & 0.450  & \textbf{0.995} (4.728) \\
         vcs-OLS & 2.007 & 0.545 & 0.357  & 0.276  & \textbf{0.999} & 0.427  & 0.874 (2.357) \\
        vl-Lasso & 2.156 & 0.539 & 0.344  & 0.976  & 0.170  & 0.488   & - (-) \\
        vcs-Lasso & 2.001 & 0.537 & 0.337  & 0.860  & 0.585  & 0.587  & - (-) \\
         \midrule
        \multicolumn{8}{|c|}{Cube: Holdout 25\%} \\
        \midrule
         l-BTRR & \textbf{1.703} & \textbf{0.693} & \textbf{0.092} & 0.907 & 0.977 & \textbf{0.922}  & 0.945 (1.574) \\
         cs-BTRR & 1.817 & 0.657 & 0.110  & \textbf{1.000}  & 0.893  & 0.884  & 0.659 (\textbf{0.894}) \\
         vl-OLS & 2.170 & 0.534 & 0.333  & 0.329  & 0.999 & 0.491   & \textbf{0.995} (4.792)  \\
         vcs-OLS & 2.040 & 0.538 & 0.332 & 0.312 & \textbf{1.000} & 0.472  & 0.869 (2.399)  \\
         vl-Lasso & 2.178 & 0.531 & 0.332  & 0.971  & 0.170  & 0.419   & - (-) \\
         vcs-Lasso & 2.039 & 0.537 & 0.331  & 0.858  & 0.580  & 0.541  & - (-) \\
         \midrule
        \multicolumn{8}{|c|}{Cube: Holdout 50\%} \\
        \midrule
         l-BTRR & \textbf{1.706} & \textbf{0.688} & \textbf{0.100} & 0.879 & 0.986 & \textbf{0.920}  & 0.971 (1.750)  \\
         cs-BTRR & 1.841 & 0.650 & 0.111  & \textbf{1.000} & 0.928  & 0.918   & 0.674 (\textbf{0.924}) \\
         vl-OLS & 2.172 & 0.532 & 0.354 & 0.305 & \textbf{0.999} & 0.463  & \textbf{0.995} (4.790) \\
         vcs-OLS & 2.042 & 0.537 & 0.351 & 0.282 & \textbf{0.999} & 0.436  & 0.869 (2.398) \\
         vl-Lasso & 2.180 & 0.530 & 0.354  & 0.969  & 0.171  & 0.418   & - (-) \\
         vcs-Lasso & 2.042 & 0.537 & 0.350  & 0.852  & 0.575  & 0.536  & - (-) \\
        \bottomrule
    \end{tabular}
    \caption{{\bf Scheme 2 Results}. Out-of-sample predictive performance (p-RMSE and p-Corr), estimated vs. true coefficient correlation across voxels (c-RMSE), feature selection (Sens, Spec, F1), and coverage probability (interval width in parentheses) for 6 competing methods. For feature selection, metrics are shown with multiplicity correction, where joint credible intervals are used for tensor-based approaches and cluster-extent inference is used for voxel-wise OLS.}
    \label{tab:Table4}
\end{table}

\begin{table}[H]
    \centering
    \begin{tabular}{|c|c|c|c|c|c|c|c|}
        \toprule
        \multicolumn{8}{|c|}{Sphere: Holdout 25\%} \\
        \midrule
        Method & p-RMSE & p-Corr & c-RMSE & Sens & Spec & F1 & Coverage \\
        \midrule
         l-BTRR & \textbf{1.758} & \textbf{0.661} & \textbf{0.168} & 0.813 & 0.970 & \textbf{0.811}  & 0.947 (2.094) \\
         cs-BTRR & 1.849 & 0.630 & 0.192 & \textbf{0.953} & 0.884  & 0.802   & 0.707 (\textbf{1.085})  \\
         vl-OLS & 2.379 & 0.449 & 0.468 & 0.209 & 0.994 & 0.315  & \textbf{0.991} (4.796) \\
         vcs-OLS & 2.045 & 0.522 & 0.366  & 0.239  & \textbf{0.998} & 0.366  & 0.869 (2.363)  \\
         vl-Lasso & 2.388 & 0.456 & 0.479  & 0.975  & 0.112  & 0.476   & - (-) \\
        vcs-Lasso & 2.041 & 0.518 & 0.361  & 0.698  & 0.597  & 0.553   & - (-) \\
        \midrule
        \multicolumn{8}{|c|}{Sphere: Holdout 50\%} \\
        \midrule
         l-BTRR & \textbf{1.748} & \textbf{0.659} & \textbf{0.180} & 0.803 & 0.961 & \textbf{0.796}  & 0.924 (1.818) \\
         cs-BTRR & 1.866 & 0.622 & 0.195 & 0.957 & 0.877  & 0.795   & 0.697 (\textbf{1.047})  \\
         vl-OLS & 2.379 & 0.448 & 0.506 & 0.187 & 0.989 & 0.287  & \textbf{0.991} (4.796) \\
         vcs-OLS & 2.046 & 0.519 & 0.387  & 0.226  & \textbf{0.997} & 0.350  & 0.868 (2.363) \\
        vl-Lasso & 2.401 & 0.453 & 0.508  & \textbf{0.969}  & 0.099  & 0.456   & - (-) \\
        vcs-Lasso & 2.042 & 0.516 & 0.375  & 0.651  & 0.575  & 0.552  & - (-) \\
         \midrule
        \multicolumn{8}{|c|}{Cube: Holdout 25\%} \\
        \midrule
         l-BTRR & \textbf{1.717} & \textbf{0.678} & \textbf{0.117} & 0.841 & 0.970 & \textbf{0.852}  & 0.950 (1.562) \\
         cs-BTRR & 1.834 & 0.636 & 0.174  & \textbf{0.971}  & 0.867  & 0.808  & 0.655 (\textbf{0.898}) \\
         vl-OLS & 2.383 & 0.438 & 0.437  & 0.205  & 0.993 & 0.309   & \textbf{0.991} (4.816)  \\
         vcs-OLS & 2.062 & 0.506 & 0.369 & 0.231 & \textbf{0.998} & 0.358  & 0.860 (2.373)  \\
         vl-Lasso & 2.397 & 0.431 & 0.438  & 0.960  & 0.109  & 0.449   & - (-) \\
         vcs-Lasso & 2.061 & 0.506 & 0.366  & 0.643  & 0.555  & 0.528  & - (-) \\
         \midrule
        \multicolumn{8}{|c|}{Cube: Holdout 50\%} \\
        \midrule
         l-BTRR & \textbf{1.719} & \textbf{0.678} & \textbf{0.130} & 0.837 & 0.959 & \textbf{0.840}  & 0.924 (1.615)  \\
         cs-BTRR & 1.860 & 0.628 & 0.176  & \textbf{0.975} & 0.953  & 0.796   & 0.643 (0.872) \\
         vl-OLS & 2.389 & 0.438 & 0.516 & 0.179 & 0.987 & 0.275  & \textbf{0.991} (4.817) \\
         vcs-OLS & 2.070 & 0.506 & 0.389 & 0.218 & \textbf{0.997} & 0.341  & 0.861 (2.373) \\
         vl-Lasso & 2.404 & 0.429 & 0.518 & 0.952  & 0.095  & 0.427   & - (-) \\
         vcs-Lasso & 2.069 & 0.506 & 0.387  & 0.632  & 0.544  & 0.523  & - (-) \\
        \bottomrule
    \end{tabular}
        \caption{{\bf Scheme 3 Results}. Out-of-sample predictive performance (p-RMSE and p-Corr), estimated vs. true coefficient correlation across voxels (c-RMSE), feature selection (Sens, Spec, F1), and coverage probability (interval width in parentheses) for 6 competing methods. For feature selection, metrics are shown with multiplicity correction, where joint credible intervals are used for tensor-based approaches and cluster-extent inference is used for voxel-wise OLS.}
    \label{tab:Table5}
\end{table}

\begin{table}[H]
    \centering
    \begin{tabular}{|c|c|c|c|c|}
        \toprule
        \multicolumn{5}{|c|}{Sphere: Holdout 25\%} \\
        \midrule
        Method & c-RMSE & Sens & Spec & F1 \\
        \midrule
        l-BTRR & \textbf{0.256} & 0.932 & 0.795 & \textbf{0.857} \\
        cs-BTRR & 0.279 & 0.899 & 0.835 & 0.847 \\
        vl-OLS & 0.367 & 0.252 & \textbf{0.999} & 0.384 \\
        vcs-OLS & 0.366 & 0.258 & \textbf{0.999} & 0.392 \\
        vl-Lasso & 0.368 & \textbf{0.977} & 0.151 & 0.380 \\
        vcs-Lasso & 0.366 & 0.834 & 0.586 & 0.590 \\
        \midrule
        \multicolumn{5}{|c|}{Sphere: Holdout 50\%} \\
        \midrule
        l-BTRR & \textbf{0.259} & 0.919 & 0.800 & \textbf{0.850} \\
        cs-BTRR & 0.280 & 0.889 & 0.814 & 0.831 \\
        vl-OLS & 0.386 & 0.226 & \textbf{0.999} & 0.353 \\
        vcs-OLS & 0.386 & 0.239 & \textbf{0.999} & 0.369 \\
        vl-Lasso & 0.388 & \textbf{0.958} & 0.153 & 0.377 \\
        vcs-Lasso & 0.387 & 0.819 & 0.579 & 0.556 \\
        \midrule
        \multicolumn{5}{|c|}{Cube: Holdout 25\%} \\
        \midrule
        l-BTRR & \textbf{0.120} & 0.941 & 0.845 & \textbf{0.865} \\
        cs-BTRR & 0.127 & 0.845 & 0.926 & 0.825 \\
        vl-OLS & 0.345 & 0.279 & \textbf{0.999} & 0.423 \\
        vcs-OLS & 0.345 & 0.300 & \textbf{0.999} & 0.445 \\
        vl-Lasso & 0.349 & \textbf{0.970} & 0.168 & 0.420 \\
        vcs-Lasso & 0.344 & 0.846 & 0.557 & 0.539 \\
        \midrule
         \multicolumn{5}{|c|}{Cube: Holdout 50\%} \\
        \midrule
        l-BTRR & \textbf{0.121} & 0.918 & 0.845 & \textbf{0.846} \\
        cs-BTRR & 0.130 & 0.804 & 0.957 & 0.802 \\
        vl-OLS & 0.364 & 0.253 & \textbf{0.999} & 0.390 \\
        vcs-OLS & 0.363 & 0.271 & \textbf{0.999} & 0.411 \\
        vl-Lasso & 0.367 & \textbf{0.961} & 0.161 & 0.385 \\
        vcs-Lasso & 0.363 & 0.830 & 0.549 & 0.519 \\
        \bottomrule
    \end{tabular}
    \caption{Simulation results for estimation with a thin strip around signal boundaries for simulation schemes 2 and 3.}\label{tab:Table6}
\end{table}

\end{document}